\documentclass[11pt]{article}
\usepackage[colorlinks=true,linkcolor=BlueViolet,citecolor=BlueViolet,urlcolor=BlueViolet]{hyperref}
\usepackage{scalerel}
\usepackage{mathtools,slashed}
\usepackage[normalem]{ulem}
\usepackage{bm}
\usepackage{cite}
\usepackage{mathrsfs}
\usepackage{amsmath}
\usepackage{graphicx}
\usepackage{verbatim}
\usepackage{color}
\usepackage[dvipsnames]{xcolor}
\usepackage{enumerate}
\usepackage{amsfonts}
\usepackage{amssymb}
\usepackage[hang,flushmargin]{footmisc}
\usepackage[font=footnotesize]{caption}
\DeclareMathAlphabet\mathbfcal{OMS}{cmsy}{b}{n}
\usepackage{empheq}
\usepackage{cancel}
\usepackage{tikz-feynman}
\usepackage{tikz}
\usepackage{pgfplots}
\pgfplotsset{compat=1.16}
\usetikzlibrary{patterns}

\csname @addtoreset\endcsname{equation}{section}

 \newcommand{\badat}{\begin{alignedat}}
 \newcommand{\eadat}{\end{alignedat}}

\def\5{\bar }
\def\6{\partial }
\def\7{\hat }
\def\4{\tilde }

\renewcommand{\d}{\partial}

\def\cA{\mathcal{A}}

\def\cD{\mathcal{D}}

\def\cH{\mathcal{H}}

\def\cJ{\mathcal{J}}

\def\cL{\mathcal{L}}

\def\cN{\mathcal{N}}
\def\cO{\mathcal{O}}

\def\cQ{\mathcal{Q}}

\definecolor{blue-violet}{rgb}{0.54, 0.17, 0.89}
\definecolor{PineGreen}{cmyk}{0.92, 0, 0.59, 0.25}
\definecolor{OliveGreen}{cmyk}{0.64, 0, 0.95, 0.40}
\definecolor{RawSienna}{cmyk}{0, 0.72, 1, 0.45}
\definecolor{Gray}{cmyk}{0, 0, 0, 0.50}
\definecolor{MidnightBlue}{cmyk}{0.98, 0.13, 0, 0.43}
\definecolor{Orange}{cmyk}{0, 0.61, 0.87, 0}
\definecolor{LimeGreen}{cmyk}{0.50, 0, 1, 0}
\definecolor{Green}{cmyk}{1, 0, 1, 0}

\renewcommand{\tilde}{\widetilde}
\renewcommand{\hat}{\widehat}
\def\beq{\begin{eqnarray}}
\def\eeq{\end{eqnarray}}
\def\beann{\begin{eqnarray*}}
\def\eeann{\end{eqnarray*}}
\def\be{\begin{equation}}
\def\ee{\end{equation}}
\def\ba{\begin{array}}
\def\ea{\end{array}}
\def\ben{\begin{enumerate}}
\def\een{\end{enumerate}}
\def\bea{\begin{eqnarray}}
\def\eea{\end{eqnarray}}

\def\tr{{\rm tr}}

\def\dI{\mathscr{I}}

\hypersetup{final}

\begin{document}
\vspace{30pt}

\begin{center}


{\Large\sc Logarithmic scalar soft factorization \\[10pt]  from matching conditions }

\vspace{-5pt}
\par\noindent\rule{350pt}{0.4pt}


\vspace{20pt}
{\sc 
Matías Briceño\quad 
Hernán A. González\quad 
Oriana Labrin
}

\vspace{10pt}

{\it\small 
Facultad de Ingeniería, Universidad San Sebastián, Santiago 8420524, Chile.}

\vspace{10pt}

{\tt\small 
\href{mailto:mbricenoc3@correo.uss.cl}{mbricenoc3@correo.uss.cl},
\href{mailto:hernan.gonzalez@uss.cl}{hernan.gonzalez@uss.cl},
\href{mailto:olabrinz@correo.uss.cl}{olabrinz@correo.uss.cl}
}
\end{center}
\vspace{20pt}

\begin{abstract}
Logarithmic soft theorems in QED and gravity can be understood as conservation laws associated with asymptotic symmetries. In contrast, for a massless scalar, no gauge symmetry is available to organize this relation. Despite that, we show that the logarithmic soft factorization for a massless scalar coupled to massive fermions through a Yukawa interaction follows from the asymptotic field equations and matching conditions at spatial infinity. From the first subleading logarithmic fall-off, we construct a conserved surface integral whose boundary values satisfy the antipodal map between past and future null infinity. At future null infinity, the field equations relate this quantity to the $1/u$ tail of the scalar radiation, while at past null infinity, for scattering states without incoming scalar radiation, it is carried by a non-radiative logarithm of the incoming particles. We derive the classical tail from the particle trajectories and obtain the hard charge from the asymptotic Dirac current. The particular solution of the leading scalar field gives the classical contribution, while its homogeneous component gives the quantum contribution. The resulting Ward identity reproduces the logarithmic soft coefficient of the one-loop scalar emission (up to a zero mode on the sphere), including its classical and quantum parts.
\end{abstract}
\newpage
\tableofcontents
\newpage

\section{Introduction}

The behavior of scattering amplitudes at low energies is tied to the asymptotic behavior of massless fields near the boundary of spacetime. In gravity and gauge theories, this relation is usually organized as an infrared triangle connecting soft theorems, memory effects, and asymptotic symmetries. The soft theorems are Ward identities of asymptotic charges \cite{Strominger:2017zoo}, and the same asymptotic degrees of freedom control classical memory observables \cite{Bieri:2013hqa, Zeldovich:1974gvh,Pasterski:2015tva,Pasterski:2015zua,Herrmann:2026kue,Diaz:2026yef}.

Here we would like to continue exploring the idea that the physics at large distances and its relation to low energies does not, in principle, require an asymptotic large transformation. This raises the question of which parts of the infrared triangle survive in a theory without a gauge symmetry to organize them. Massless scalar fields coupled to massive matter provide a natural setting for this question. Even without shift symmetry\footnote{Goldstone modes provide an interesting laboratory for studying the infrared physics in effective field theories. In this case, scalar soft theorems take a geometric form, in which the zero-energy limit acts as a covariant derivative on the field space of vacua \cite{Cheung:2021yog,Derda:2024jvo}. At one loop, this form is preserved for derivative couplings and corrected by potential interactions \cite{Cohen:2025dex}. }, such scalars admit a leading soft theorem and a family of asymptotic charges \cite{Campiglia:2017dpg}. A possible interpretation of these charges in terms of a dual two-form field has also been discussed \cite{Francia:2018jtb,Henneaux:2018mgn,Campiglia:2018see,Romoli:2024hlc,Romoli:2025map}.

Beyond tree level, the asymptotic analysis must include logarithms, as shown in \cite{Fuentealba:2024lll}. In four dimensions, long-range forces modify the asymptotic motion of massive particles: their trajectories acquire logarithmic corrections and the radiated field is accompanied by $1/u$ tails. The relation between these tails and the $\log\omega$ terms of the soft expansion was established for classical and quantum scattering in \cite{Sahoo:2018lxl}, and for scalar radiation a direct worldline calculation determines the late-time waveform  \cite{Duary:2025siq}.

Matching conditions across spatial infinity \cite{Strominger:2013jfa,Fuentealba:2023syb,Capone:2022gme,Fuentealba:2025ekj,Compere:2025bnf,Compere:2026jmk} have also been used to derive classical and quantum logarithmic soft theorems in electrodynamics and gravity \cite{Campiglia:2019wxe,Donnay:2020lur,Choi:2024mac,Choi:2024ygx,Choi:2024ajz,Boschetti:2025tru,Compere:2026vvi,Boschetti:2026gfd}, to construct subleading charges \cite{Kapec:2014opa,Lysov:2014csa,Campiglia:2014yka,Campiglia:2016hvg,Nagy:2024jua,Barnich:2024vks,Briceno:2025ivl,Flanagan:2026yzu,Charbonnier:2026vlu}, and to relate them to physical fluxes at null infinity \cite{Briceno:2025cdu}. 

The behavior of the free scalar near spatial infinity is the starting point of our construction. The scalar equation near $i^0$ admits two independent branches of solutions, and one of them (the $Q$-branch) produces logarithms near null infinity \cite{Henneaux:2018mgn,Fuentealba:2024lll}. The analysis presented in \cite{Fuentealba:2024lll} showed that these branches obey a logarithmic matching between past and future null infinity. Here we use the first subleading logarithmic sector to build a conserved surface integral on the spatial hyperboloid,  whose boundary values give charges at $\dI^+_-$ and $\dI^-_+$ that are antipodally identified on the celestial sphere. 

Relative to the free wave equation analysis at $i^{0}$ of \cite{Henneaux:2018mgn,Fuentealba:2024lll}, we couple the scalar to massive matter and identify the physical content of the logarithmic charge: its hard part is determined by the logarithmic source at timelike infinity, and it is directly measured by the $1/u$ tail of the radiation. In this sense, the soft part of the charge is carried by the scalar radiation,
its hard part is fixed by the massive particles, and the conservation law follows from asymptotic field equations and matching alone. Relative to \cite{Campiglia:2017dpg,Briceno:2025ivl}, which are tree-level statements, the charge constructed in this work is sensitive to one-loop physics through the homogeneous component of the scalar field. Relative to \cite{Duary:2025siq}, we identify the classical tail as the classical part of a conserved charge. Finally, we find that for physical scattering states, the logarithmic data at past null infinity are non-radiative: the incoming particles produce a non-radiative term $\log r/r^2$ at $\dI^-$, without a $1/v$ tail, and this term is precisely the one required by the antipodal matching with the $1/u$ tail at $\dI^+_-$. This is consistent with the analysis presented in \cite{AtulBhatkar:2019vcb,Compere:2025tzr} for electrodynamics and \cite{Compere:2026vvi} for gravity.

In our construction, the main model is a massless real scalar coupled to a massive Dirac field through a Yukawa interaction, with the massive scalar matter treated in parallel. We first establish the classical particle description and determine its radiation tail. We then derive the logarithmic source near timelike infinity and propagate it to null infinity. The resulting hard charge depends on the leading scalar field near timelike infinity, which satisfies an inhomogeneous differential equation on the hyperboloid. Its particular solution reproduces the projection on the sphere of the classical radiation tail, while the matrix elements of its homogeneous solution, evaluated using the leading scalar soft relation, involve the logarithmic function of the hard momenta that also appears in one-loop infrared divergences. Charge conservation then yields a Ward identity, which we verify by matching the projected logarithmic coefficient of the one-loop soft emission amplitude. This construction parallels the treatment of the loop-corrected soft photon theorem presented in \cite{Campiglia:2019wxe,AtulBhatkar:2019vcb}.

The paper is organized as follows. In Section~\ref{sec:setup}, we fix our conventions. In Section~\ref{sec:first}, we rederive the corrected classical trajectories, the scalar tails, and the leading memory. In Section~\ref{sec:charges}, we construct the logarithmic charge, derive its matching and relate its soft and hard decomposition to tail memory. Section~\ref{sec:interacting} contains the asymptotic analysis of the interacting theories near timelike infinity, and Section~\ref{sec:retarded} propagates the logarithmic source to null infinity. The action of the hard charge, with its classical and homogeneous contributions, is computed in Section~\ref{sec:hard}. Section~\ref{sec:oneloop} contains the one-loop soft emission, and Section~\ref{sec:Ward} the Ward identity. We discuss the results in Section~\ref{sec:discussion}.  The appendices contain the evaluation of the one-loop integrals, the renormalization of Yukawa theory, and non-radiative logarithms at past null infinity together with their antipodal matching.

\section{Setup and conventions}
\label{sec:setup}

We work in four-dimensional Minkowski spacetime with mostly-plus signature. The radiated field is a real massless scalar $\phi$, and the massive matter is a Dirac spinor $\Psi$. The renormalized Yukawa Lagrangian is
\begin{equation}
\cL=-\frac{1}{2}(\d\phi)^2
+\frac{i}{2}\Big(\bar\Psi\gamma^\mu\d_\mu\Psi-\d_\mu\bar\Psi\,\gamma^\mu\Psi\Big)
-M\bar\Psi\Psi+g\,\phi\,\bar\Psi\Psi+\cL_{\rm ct}\,,
\label{YukawaL}
\end{equation}
with
\begin{equation}
\{\gamma^\mu,\gamma^\nu\}=-2\eta^{\mu\nu}\,,\qquad
\bar\Psi=\Psi^\dagger\gamma^0\,,\qquad
\slashed p=\gamma^\mu p_\mu\,,
\end{equation}
so that $(\gamma^0)^2=1$ and $\slashed p^{\,2}=-p^2$. The classical field equations, before taking into account counterterms, are
\begin{equation}
(i\slashed\d-M+g\phi)\Psi=0\,,\qquad \Box\phi=-g\,\bar\Psi\Psi\,.
\label{YukawaEOM}
\end{equation}
For an on-shell external spinor, $\bar u(p)(-\slashed p+M)=2M\,\bar u(p)$, and we normalize them using $\bar u_s u_{s'}=2M\delta_{ss'}$. All momenta $p_i^\mu$ are future directed, and the soft momentum is parametrized by
\begin{equation}
p_i^2=-M^2\,,\qquad
k^\mu=\omega q^\mu\,,\qquad
q^\mu(\hat{x})=(1,\hat x)\,,
\label{kinematics}
\end{equation}
with $\eta_i=1$ ($\eta_i=-1$) for outgoing (incoming) particles. The symbol $q^\mu$ always denotes the soft null direction. We denote loop momenta by $\ell^\mu$, and the scalar coupling of a classical particle by $\kappa_a$.

\noindent With these conventions, the leading soft relation reads
\begin{equation}
\cA_{N+1}(\omega q;\{p_i\})=\frac{1}{\omega}\left(gM\sum_i\frac{\eta_i}{p_i\cdot q} \right)\,\cA_N(\{p_i\})+O(\omega^0)\,.
\label{leadingSoft}
\end{equation}
The above soft expansion contains subleading corrections. In order to extract the coefficient of $\log\omega$, we will use the operator
\begin{equation}
L_\omega[F]=\lim_{\omega\to0^+}\omega\,\d_\omega^2(\omega F)\,,\qquad
L_\omega\!\left[\frac{f_{-1}}{\omega}+f_{\log}\log\frac{\omega}{\omega_0}+f_0\right]=f_{\log}\,,
\label{Lomega}
\end{equation}
which removes the leading soft pole and the constant term, thus
isolating the logarithmic coefficient $f_{\log}$. Here $\omega_0$ is an arbitrary energy scale introduced to make the argument of the logarithm dimensionless. 
\paragraph{Renormalization conditions and local scalar interactions.}
For the one-loop computations, we primarily focus on the Yukawa theory as the coupling $g$ is dimensionless. We impose renormalization conditions to keep a vanishing massless scalar pole with residue equal to one,
\begin{equation}
\Pi_\phi^{\rm ren}(0)=0\,,\qquad
\left.\frac{d\Pi_\phi^{\rm ren}(k^2)}{dk^2}\right|_{k^2=0}=0\,.
\label{renCond}
\end{equation}
 Masslessness of the field is imposed by hand, due to the absence of symmetries that would protect it at loop level. Since the fermion mass breaks the discrete symmetry $\phi\to-\phi$, $\Psi\to\gamma_5\Psi$, renormalization also requires local counterterms for the scalar tadpole and for $\phi^3$ and $\phi^4$. As in \cite{Biswas:2022lsj}, we choose the counterterms so that, order by order, these self-interactions are canceled, and only the interactions present at tree level remain; in particular, the renormalized cubic and quartic couplings vanish at tree level. In \cite{Biswas:2022lsj} the existence of such counterterms is assumed. Here we compute explicitly the one-loop scalar mass and wave-function counterterms fixed by \eqref{renCond}, together with the fermion and Yukawa-vertex counterterms (Appendix~\ref{AppRen}). With this choice, scalar self-interactions do not enter the comparison made in Section~\ref{sec:oneloop} at order $g^3$.
\section{Worldline action driven by a massless scalar}
\label{sec:first}

We start by revisiting the classical scalar radiation analysis of \cite{Duary:2025siq}. Our aim is to characterize the long-range corrections associated to the scalar waveform, which will later be identified with the classical contribution to a conservation law.

We consider a scattering process involving several massive particles of masses $m_a$ interacting through a massless scalar field $\Phi(x)$. At leading order, the particle trajectories are asymptotically moving at constant velocity. The long-range scalar interaction, however, induces logarithmic corrections to their asymptotic motion and produces scalar radiation that reaches null infinity. The dynamics is described by
\begin{equation}
S[\Phi, X_a] \;=\;  \int d^4x \;
\Big(-\tfrac{1}{2}(\partial \Phi )^2 +\Phi J \Big)
+ \sum_a \frac{m_a}{2} \int d\tau_a \, (\dot{X}_a(\tau_a))^2\,,
\qquad
J(x)= \sum_a \kappa_a \int d\tau_a \,  \delta^{(4)} (x-X_a(\tau_a))\,,
\label{ppAction}
\end{equation}
where $\kappa_a$ is the scalar charge of particle $a$. The equations of motion are
\begin{align}
    \Box\Phi(x)={}& -J(x) \,,
    \\ m_a\ddot{X}^\mu_a(\tau_a)={}&  \kappa_a\partial^\mu\Phi(X_a(\tau_a))\,.
\end{align}
The parameter $\tau_a$ is not assumed to be the proper time along the interacting trajectory. It increases along the worldline and is normalized so that the asymptotic velocity has unit norm. The incoming and outgoing asymptotic regions correspond to $\tau_a\to-\infty$ and $\tau_a\to+\infty$, and we label them by $\sigma=-1$ and $\sigma=+1$, respectively.

Each particle contributes to the field, $\Phi=\sum_b\Phi_b$. Here we consistently neglect the interaction of the field $a$ in its own equation of motion. In this way, the equations relevant for its asymptotic motion are
\begin{align}   \Box\Phi_b(x)={}& -J_b(x) \,,\label{Eom1}
    \\ m_a\ddot{X}^\mu_a(\tau_a)={}& \sum_{b\neq a} \kappa_a\partial^\mu\Phi_b(X_a(\tau_a))\,\label{Eom2}.
\end{align}
In what follows, we study these equations perturbatively always using the retarded propagation.  

\subsection{Logarithmic corrections to the trajectories}

In flat space, the retarded field is given by
\begin{equation}
    \Phi_{\rm R}(x) = \sum_a \frac{\kappa_a }{2\pi}\int d\tau_a \, \theta(x^0 - X^0_a(\tau_a))\, \delta\big(-(x-X_a(\tau_a))^2\big)
    =\sum_a\frac{\kappa_a}{4\pi}\frac{1}{|\dot X_a\cdot(x-X_a)|}\bigg|_{\tau_a=\tau_a^*}\,,
       \label{phi}
\end{equation}
where $\tau_a^*$ is the retarded root of $(x-X_a(\tau_a))^2=0$. At leading order in the asymptotic regions, the trajectories are moving at constant velocity,
\begin{equation}
X_a^\mu(\tau_a)
=
\alpha_{a,\sigma}^\mu
+
V_{a,\sigma}^\mu\tau_a
+\cdots,
\qquad
\tau_a\rightarrow\sigma\infty\,,
\qquad
V_{a,\sigma}^\mu=\frac{p_{a,\sigma}^\mu}{m_a}\,,\qquad V_{a,\sigma}^2=-1\,,
\label{leadingtrajectory}
\end{equation}
where $\alpha_{a,\sigma}^{\mu}$ is a constant asymptotic displacement. For a linear trajectory, the null condition gives
\begin{equation}
\tau_a^*
=
-V_{a,\sigma}\cdot(x-\alpha_{a,\sigma})
\pm
\sqrt{
\bigl[
V_{a,\sigma}\cdot(x-\alpha_{a,\sigma})
\bigr]^2
+
(x-\alpha_{a,\sigma})^2
}\,,
\end{equation}
with the sign fixed by $x^0>X_a^0(\tau_a^*)$, and the gradient of the field generated by particle $b$ is
\begin{equation}
\partial^\mu\Phi_b(x)
=
-\frac{\kappa_b}{4\pi}
\frac{
\bigl[
V_{b,\sigma}\cdot(x-\alpha_{b,\sigma})
\bigr]V_{b,\sigma}^\mu
+
(x-\alpha_{b,\sigma})^\mu
}{
\left\{
\bigl[
V_{b,\sigma}\cdot(x-\alpha_{b,\sigma})
\bigr]^2
+
(x-\alpha_{b,\sigma})^2
\right\}^{3/2}
}\,.
\label{gradientScalar}
\end{equation}
We evaluate it on the leading trajectory of particle $a$. Writing $\alpha_{ab,\sigma}=\alpha_{a,\sigma}-\alpha_{b,\sigma}$, the displacement is suppressed by $1/|\tau_a|$. The numerator of \eqref{gradientScalar} is proportional to $\tau_a$ and the denominator to $|\tau_a|^3$, and since $\operatorname{sgn}(\tau_a)=\sigma$,
\begin{equation}
\partial^\mu\Phi_b(X_a)
=
-\sigma\,
\frac{\kappa_b}{4\pi\tau_a^2}
\frac{
z_{ab,\sigma}
V_{b,\sigma}^\mu
+
V_{a,\sigma}^\mu
}{
w_{ab,\sigma}^3
}
+
O\!\left(\frac{\log|\tau_a|}{|\tau_a|^{3}}\right),
\qquad
z_{ab,\sigma}=V_{a,\sigma}\cdot V_{b,\sigma}\,,\quad
w_{ab,\sigma}=\sqrt{z_{ab,\sigma}^2-1}\,.
\label{asymptoticForce}
\end{equation}
We assume non-vanishing relative velocities between distinct asymptotic particles. The interesting point is that the order $1/\tau_a^2$ in the  force implies a logarithmic correction to the asymptotic trajectories,
\begin{equation}
X_a^\mu(\tau_a)
=
\alpha_{a,\sigma}^\mu
+
V_{a,\sigma}^\mu\tau_a
+
C_{a,\sigma}^\mu\log\frac{|\tau_a|}{\tau_0}
+O\!\left(\frac{\log|\tau_a|}{|\tau_a|}\right),
\qquad
\tau_a\rightarrow\sigma\infty\,.
\label{correctedtrajectory}
\end{equation}
In the above, the reference scale $\tau_0$ only shifts $\alpha^\mu_{a,\sigma}$ and leaves $C^\mu_{a,\sigma}$ unchanged. Since $\ddot X_a^\mu=-C_{a,\sigma}^\mu/\tau_a^2+\cdots$, plugging \eqref{asymptoticForce} back into \eqref{Eom2} gives
\begin{equation}
C_{a,\sigma}^\mu
=
\sigma
\sum_{\substack{b\in\mathcal P_\sigma\\ b\neq a}}
\frac{\kappa_a\kappa_b}{4\pi m_a}
\frac{
z_{ab,\sigma}
V_{b,\sigma}^\mu
+
V_{a,\sigma}^\mu
}{
w_{ab,\sigma}^3
}\,,
\label{Ccl}
\end{equation}
where $\mathcal P_\sigma$ denotes the set of particles
in the asymptotic sector $\sigma$, with $\mathcal P_-$
and $\mathcal P_+$ corresponding to incoming and outgoing
particles, respectively. The sum excludes self-interactions
and includes only particles belonging to the same sector. 

\subsection{Retarded waveform}

The logarithmic correction to the particle trajectories produces a $1/u$ tail in the retarded waveform at future null infinity.
We use retarded time $u=t-r$ and the null parameterization $q^\mu=(1,\hat x)$.
The early- and late-time limits probe the incoming ($\sigma=-1$) and outgoing ($\sigma=+1$) trajectories, respectively.
Solving $q\cdot X_a(\tau_a^*)=-u$ for the retarded root at large $r$ and expanding \eqref{phi} at large $|u|$ gives
\begin{equation}
\Phi_R(u,r,\hat x)
=
\frac{\varphi_R(u,\hat x)}{r}+O(r^{-2})\,,
\qquad
\varphi_R(u,\hat x)
=
\varphi^+_\sigma(\hat x)+\frac{\chi^{+,\rm pp}_\sigma(\hat x)}{u}
+O\!\left(\frac{\log|u|}{u^2}\right),
\label{retardedWaveform}
\end{equation}
with
\begin{equation}
\varphi^+_\sigma(\hat x)
=
-\frac{1}{4\pi}\sum_{{a\in\mathcal P_\sigma}}
\frac{\kappa_a}{q\cdot V_{a,\sigma}}\,,
\qquad
\chi^{+,\rm pp}_\sigma(\hat x)
=
-\frac{1}{4\pi}\sum_{{a\in\mathcal P_\sigma}}
\frac{\kappa_a\,q\cdot C_{a,\sigma}}{q\cdot V_{a,\sigma}}\,.
\label{retardedEndpointCoefficients}
\end{equation}
The coefficient $\varphi^+_\sigma$ is the field of the asymptotic straight
trajectories. On the other hand, let us keep in mind that the dependence of $\chi^{+,\rm pp}_\sigma$ on the
asymptotic momenta will reappear in the construction of the surface integral of Section~\ref{sec:charges}, specifically in its hard contribution.

At $\dI^-$ we use advanced time $v=t+r$ and
$\bar q^\mu=(-1,\hat x)$, with $\bar q^\mu(-\hat x)=-q^\mu(\hat x)$.
At fixed $v$, the retarded root tends to the past as
$-2r\,\bar q\cdot V_{a,-}+O(\log r)$.
The leading $1/r$ coefficient depends only on the incoming velocities
and is independent of $v$. The trajectory correction produces an $r^{-2}\log r$ term, whose role in the antipodal matching we also discuss in Section~\ref{sec:charges} and compute in Appendix~\ref{AppCoulomb}.

\subsubsection*{Angular form of the tail} 
For later comparison with the hard charge in Section~\ref{sec:hard}, we rewrite \eqref{retardedEndpointCoefficients} in a form adapted to the hyperboloid arising at time-like infinity. At $i^{+}$, we set $V^\mu_{a,+}=Y^\mu_a$ and define
\begin{equation}
\cD_i(q)f=\frac{q_\mu\nabla^AY_i^\mu}{q\cdot Y_i}\,\nabla^{(i)}_Af\,,
\qquad
\nabla_AY^\mu\nabla^AY^\nu=\eta^{\mu\nu}+Y^\mu Y^\nu\,,
\label{Tdef}
\end{equation}
where $\nabla_A$ is the covariant derivative on the unit hyperboloid and the label $(i)$ indicates which argument is differentiated. Using $\nabla_A(Y_a\cdot Y_b)=\nabla_AY_a\cdot Y_b$ and the last relation in \eqref{Tdef},
\begin{equation}
\cD_a(q)\left[\frac{1}{\sqrt{(Y_a\cdot Y_b)^2-1}}\right]
=-\frac{q\cdot Y_a+(Y_a\cdot Y_b)\,q\cdot Y_b}{(q\cdot Y_a)\big[(Y_a\cdot Y_b)^2-1\big]^{3/2}}-\frac{1}{\sqrt{(Y_a\cdot Y_b)^2-1}}\,.
\label{TK}
\end{equation}
Comparing with \eqref{Ccl}\footnote{For a single species, $\kappa_a=g$ and $m_a=M$, as follows from matching the worldline soft factor $\kappa_a/(V_a\cdot k)$ with the Yukawa factor $gM/(p_a\cdot k)$ in \eqref{leadingSoft}} the outgoing tail therefore satisfies
\begin{equation}
D^2\chi^{+,\rm pp}_+=\frac{g^3}{16\pi^2M}\,D^2\sum_{\substack{{a,b\in\mathcal P_+}\\a\neq b}}\cD_a(q)\left[\frac{1}{\sqrt{(Y_a\cdot Y_b)^2-1}}\right]\,,
\label{projectedParticleTail}
\end{equation}
where $D^2$ is the Laplacian on the sphere of directions $\hat x$. The second term in \eqref{TK} does not depend on $\hat x$ and is removed by $D^2$. This is the
classical quantity that we compare with the hard charge in Section~6.

\subsection{Leading scalar memory}
\label{sec:memory}
Leading scalar memory measures the change in the leading $1/r$
field between the incoming and outgoing configurations
\cite{Campiglia:2017dpg,Duary:2025siq}. We denote the final and initial
values of this coefficient by $\varphi^\pm_\sigma$, where the
superscript labels null infinity and $\sigma$ labels its early or late end. Remember that all four values always refer to the same retarded field.

Let $\varphi_R^-(v,\hat x)$ denote the leading coefficient of the retarded field at past null infinity. Although leading memory
can be defined from the change of the field along $\dI^+$,
we include both null boundaries to express it in terms of the
incoming and outgoing asymptotic data.
The leading antipodal
matching identifies the two values adjacent to spatial infinity,
$\varphi^+_-(\hat x)=\varphi^-_+(-\hat x)$
\cite{Campiglia:2017dpg,Henneaux:2018mgn}.

We choose the integration directions toward these corners: from $u=+\infty$ to $u=-\infty$ on $\dI^+$, and from
$v=-\infty$ to $v=+\infty$ on $\dI^-$. Defining $\Delta\varphi$ as the future oriented integral minus the past
oriented integral, we obtain
\begin{equation}
    \begin{aligned}
    \Delta\varphi(\hat x)
    &=
    \left[-\int_{-\infty}^{+\infty}du\,
    \partial_u\varphi_R(u,\hat x)\right]
    -\int_{-\infty}^{+\infty}dv\,
    \partial_v\varphi_R^-(v,-\hat x)\\
    &=
    \varphi^-_-(-\hat x)-\varphi^+_+(\hat x)\,.
    \end{aligned}
\label{memoryallterms}
\end{equation}
For the retarded scattering solution, the leading coefficient at
$\dI^-$ is independent of $v$ by \eqref{coulombLog}. Its early and
late values therefore agree. The late value is matched to the
early value at $\dI^+$, so
\[
    \varphi^-_-(-\hat x)=\varphi^-_+(-\hat x)=\varphi^+_-(\hat x)\,.
\]
Using this identity in \eqref{memoryallterms} and substituting
\eqref{retardedEndpointCoefficients}, we obtain
\begin{equation}
\Delta\varphi(\hat x)=-\frac{1}{4\pi}
    \left[
    \sum_{a\in{\rm in}}
       \frac{\kappa_a}
       {q\cdot V_{a,-}}   -
        \sum_{a\in{\rm out}}
        \frac{\kappa_a}
        {q\cdot V_{a,+}}
    \right]
    \,.
\label{scalarMemoryParticles}
\end{equation}
Section~\ref{sec:tailmemory} considers the corresponding change in the $1/u$ tail.

\section{Charges construction}
\label{sec:charges}

The absence of a gauge symmetry makes the notion of charge less direct for a scalar field than in gauge theories. Inspired by the spatial infinity analysis of \cite{Fuentealba:2024lll}, we construct a surface integral whose conservation follows directly from the asymptotic equations of motion near spatial infinity.

\subsection{The field near spatial infinity}
\label{sec:spatial-infinity-charge}

We begin by determining the form of the field near spatial infinity. Since massive particles approach time-like rather than spatial infinity, the scalar satisfies the free wave equation in a neighborhood of $i^0$. We use hyperbolic coordinates
\begin{equation}
    \eta=\sqrt{r^2-t^2}\,,
    \qquad
    s=\frac{t}{r}\,,
\end{equation}
in which $\eta\to\infty$ at fixed $s\in(-1,1)$ approaches $i^0$, while $s\to1$ and $s\to-1$ approach the past boundary of future null infinity, $\dI^+_-$, and the future boundary of past null infinity, $\dI^-_+$, respectively. For a free field the expansion \cite{Henneaux:2018mgn,Fuentealba:2024lll}
\begin{equation}
    \Phi(\eta,s,x^A)
    =
    \sum_{k=0}^{\infty}
    \eta^{-(k+1)}
    (1-s^2)^{\frac{1-k}{2}}
    \Theta^{(k)}(s,x^A)\,,
    \label{PhisolutionH}
\end{equation}
reduces the wave equation to
\begin{equation}
    (1-s^2)\partial_s^2\Theta^{(k)}
    +
    2(k-1)s\,\partial_s\Theta^{(k)}
    -
    \left[
        D^2+k(k-1)
    \right]
    \Theta^{(k)}
    =
    0\,.
    \label{HyperboloidEQ}
\end{equation}
However, the scattering states require a more general expansion than the one above. The reason is that the incoming particles move along logarithmically corrected trajectories, $X^\mu=V^\mu\tau+C^\mu\log|\tau|+\cdots$, and for a point near $i^0$ their retarded root lies at $|\tau|\propto\eta$ (see  Appendix~\ref{AppCoulomb}). The field therefore depends on $\log\eta$, and the consistent expansion adds a new logarithmic contribution
\begin{equation}
    \Phi(\eta,s,x^A)
    =
    \sum_{k=0}^{\infty}
    \eta^{-(k+1)}
    (1-s^2)^{\frac{1-k}{2}}
    \Big[\Theta^{(k)}(s,x^A)+\log\eta\,\Lambda^{(k)}(s,x^A)+\cdots\Big]\,.
    \label{polyhomogeneous}
\end{equation}
In the rest of the discussion, we focus on the $k=1$ sector, where the ansatz allows both an
$\eta^{-2}\log\eta$ term and an $\eta^{-2}$ term. The logarithmic
coefficient $\Lambda^{(1)}$ satisfies a homogeneous equation and
acts as a source for $\Theta^{(1)}$, the coefficient of the term
without a logarithm. Substituting the ansatz into the free wave
equation and collecting the terms at orders $\eta^{-4}\log\eta$
and $\eta^{-4}$ gives
\begin{equation}
(1-s^2)\partial_s^2\Lambda^{(1)}-D^2\Lambda^{(1)}=0\,,\qquad
(1-s^2)\partial_s^2\Theta^{(1)}-D^2\Theta^{(1)}=-\frac{2}{1-s^2}\,\Lambda^{(1)}\,.
\label{logEtaEqs}
\end{equation}
The homogeneous
equation in \eqref{logEtaEqs} has two independent branches,
\begin{equation}
F^P_\ell(s)=(1-s^2)P'_\ell(s)\,,
\qquad
F^Q_\ell(s)=(1-s^2)Q'_\ell(s)\,,
\label{ThetaQ1}
\end{equation}
where $P_\ell$ and $Q_\ell$ are Legendre functions of the first and second kind, and $\ell\geq1$ is a spherical harmonic.

In what follows, we set the coefficient of $F^Q_\ell$ to zero in  $\Lambda^{(1)}$, since it would
generate a $\log r/(ru)$ term at null infinity that would spoil the assumed radiative boundary condition \eqref{retardedWaveform}, leaving only the $P-$branch\footnote{Notice that this restriction applies only to the logarithmic coefficient since the field $\Theta^{(1)}$ may still contain homogeneous contributions from both branches. This is confirmed by Appendix \ref{AppCoulomb} where we can see that $\Lambda^{(1)}$ vanishes at both endpoints for the retarded field.}.

The equation for $\Theta^{(1)}$ has a source fixed by
$\Lambda^{(1)}$. Its solution contains a particular term
proportional to a Legendre polynomial, together with a
homogeneous contribution from each branch. In the sector
with $\ell\geq1$, the two coefficients are therefore
\begin{equation}
\begin{aligned}
\Lambda^{(1)}
&=\sum_{\ell\geq1,m}\Lambda_{\ell m}F^P_\ell(s)Y_{\ell m}\,,\\
\Theta^{(1)}
&=\sum_{\ell\geq1,m}
\Big[
\Omega_{\ell m}F^P_\ell(s)
+\Theta_{\ell m}F^Q_\ell(s)
-\Lambda_{\ell m}P_{\ell-1}(s)
\Big]Y_{\ell m}\,.
\end{aligned}
\label{ThetaLogEta}
\end{equation}
We collect the coefficients into functions on the sphere,
\begin{equation}
\Lambda(x^A)=\sum_{\ell\geq1,m}\Lambda_{\ell m}Y_{\ell m}(x^A)\,,\qquad \Theta(x^A)=\sum_{\ell\geq1,m}\Theta_{\ell m}Y_{\ell m}(x^A)\,.
\end{equation}
The coefficients $\Omega_{\ell m}$ do not produce logarithms, and they will not be relevant for the conserved quantities we are interested in. For $\Lambda=0$, \eqref{ThetaLogEta} reduces to the logarithmic branch of the analysis presented in \cite{Henneaux:2018mgn,Fuentealba:2024lll}.

In summary, at the order relevant for the $r^{-2}$ fall-off, the field near spatial infinity is determined by two functions on the sphere: the coefficient $\Theta$ of the $Q$-branch, and the coefficient $\Lambda$ of the $\log\eta$ term, which is produced by the logarithmic corrections to the trajectories. We now construct a conserved quantity from this solution.

\subsection{A conserved surface integral}
\label{sec:conserved}

Following the strategy presented in \cite{Fuentealba:2024lll}, we propose the surface integral
\begin{equation}
    \mathcal Q
    =
    \int_{S^2}
    d^2x\,\sqrt{\gamma}\,
    \Big[ \theta\partial_s\Theta^{(1)}  -\Theta^{(1)}\partial_s\theta  + \lambda\partial_s\Lambda^{(1)}-\Lambda^{(1)}\partial_s \lambda \Big]\,.
    \label{GeneralSpatialCharge}
\end{equation}
where we use a pair of smearing functions $\lambda$ and $\theta$ that satisfy the same equations than \eqref{logEtaEqs}
\be
\begin{split}
(1-s^2)\partial_s^2\theta-D^2\theta&=0\,,\\
(1-s^2)\partial_s^2\lambda-D^2\lambda
&=-\frac{2}{1-s^2}\theta\,.
\end{split}
\ee
$\mathcal{Q}$ is independent of $s$ as follows from these equations yielding
\begin{equation}
    \frac{d}{ds}\mathcal Q=0\,.
\end{equation}
In obtaining the above, we have integrated by parts on the sphere. The quantity $\mathcal Q$ is therefore constant on any $s={\rm const.}$ slice of the spatial hyperboloid on which it is evaluated. Since $s$ plays the role of a timelike coordinate on the hyperboloid, this conservation law allows us to evaluate the same quantity on any cut and, in particular, at the points $s\to\pm1$, which approach the corners $\dI^+_-$ and $\dI^-_+$ of null infinity.

We now evaluate the conserved surface integral that identifies the asymptotic scalar field component associated with the tail field at future null infinity. Since the smearing parameters satisfy the same coupled equations as $\Theta^{(1)}$ and $\Lambda^{(1)}$, we may choose them from the same space of asymptotic solutions. 
For $\ell\geq 1$, the $F^P_\ell$ vanishes at the limits $s=\pm1$, while $F^Q_\ell$ approaches finite, nonzero values. These surviving values determine the $1/u$ and $1/v$ tails of the leading radiative field at future and past null infinity, respectively. For our purposes, we associate $\theta$ with $F^P_\ell$, such that we can isolate the homogeneous $Q$-branch contribution of $\Theta^{(1)}$. 

From the previous discussion, we can express the smearing function in terms of a single function of the sphere\footnote{We use the normalization $\int_{S^2} d^2x\,\sqrt{\gamma}\,
Y_{\ell m}(x^A)Y_{\ell' m'}(x^A)
=
\delta_{\ell\ell'}\delta_{mm'}\,.
$}, 
$\mu(x^A)=\sum_{\ell,m}\mu_{\ell m}Y_{\ell m}(x^A)$, we take
\begin{equation}
 \lambda
    =
    \frac12\sum_{\ell\geq1,m}
    \mu_{\ell m}
    P_{\ell-1}(s)
    Y_{\ell m}(x^A)\,,
    \quad
    \theta
    =
    -\frac12\sum_{\ell\geq1,m}
    \mu_{\ell m}
    F_\ell^P(s)
    Y_{\ell m}(x^A)\,.
    \label{epsilonP1}
\end{equation}
By orthonormality, the integral on the sphere pairs each real spherical harmonic only with itself, so we can evaluate the pairing one mode at a time. The pairing of two $P$-branch solutions vanishes, and the particular terms proportional to $P_{\ell-1}$ in $\Theta^{(1)}$ and $\lambda$ cancel against each other. This leaves the pairing of the auxiliary $P$ branch with the field's $Q$ branch. The Legendre equation and the Wronskian identity\footnote{We use
$\partial_s[(1-s^2)P'_\ell]=-\ell(\ell+1)P_\ell$, with the
same identity for $Q_\ell$, and
$P_\ell Q'_\ell-Q_\ell P'_\ell=1/(1-s^2)$.}
then give
\begin{equation}
F_\ell^P\partial_sF_\ell^Q-F_\ell^Q\partial_sF_\ell^P
=\ell(\ell+1)\,.
\label{Wronskian}
\end{equation}
With the normalization in \eqref{epsilonP1}, the charge is expressed in terms of one single smearing function $\mu$, yielding
\begin{equation}
\mathcal Q[\mu]
=
-\frac12\sum_{\ell\geq1,m}\ell(\ell+1)\,\mu_{\ell m}\Theta_{\ell m}
=
\frac12\int_{S^2}d^2x\sqrt\gamma\,\mu\,D^2\Theta\,.
\label{fmuRelation}
\end{equation}
The constant mode of $\mu$ does not contribute, so the charge is parametrized by functions on the sphere modulo constants.

In summary, the integral \eqref{GeneralSpatialCharge} is conserved by the field equations near $i^0$ alone, and it measures the $Q$-branch coefficient $\Theta$. Being independent of $s$, it can be evaluated at both ends of the hyperboloid. To give it a physical meaning, we now express $\Theta$ in terms of the data at null infinity.

\subsubsection*{Null infinity data and matching} 

We now evaluate $\mathcal Q$ at the boundaries of the hyperboloid $s\to\pm1$ and express it in terms of the asymptotic field at the corners of null infinity. Near future null infinity we use retarded coordinates, $t=u+r$, and write the leading radial coefficient at large $|u|$ as
\begin{equation}
    \Phi=\frac{\phi_0(u,x^A)}{r}+O\!\left(\frac{\log r}{r^2}\right),\qquad
    \phi_0(u,x^A)
    =
    \phi^{(0,0)}(x^A)
    +
       \sum_{m\geq1}
    \frac{\phi^{(0,m)}(x^A)}{u^m}
    \,,
    \label{phi0LargeU}
\end{equation}
and analogously at past null infinity with advanced time $v$, $t=v-r$. The values of the tail coefficients are
\begin{equation}
\chi^+_\pm(x^A)\equiv\phi^{(0,1)}(x^A)\big|_{\dI^+_\pm}\,,
\qquad
\chi^-_\pm(x^A)\equiv\phi^{(0,1)}(x^A)\big|_{\dI^-_\pm}\,.
\label{nullTailCoefficients}
\end{equation}
At order $r^{-2}$ the field may contain both radial and temporal
logarithms. We extract the coefficient of $r^{-2}\log r$ at fixed
null time before taking the endpoint limits,
\be
\Phi^+_-(x^A)
=\lim_{u\to-\infty}\operatorname{Coeff}_{r^{-2}\log r}
  \bigl[\Phi(u,r,x^A)\bigr]\,, \quad
\Phi^-_+(x^A)
=\lim_{v\to+\infty}\operatorname{Coeff}_{r^{-2}\log r}
  \bigl[\Phi(v,r,x^A)\bigr]\,.
\ee
The matching across spatial infinity relates two coefficients at each corner, the tail and the radial logarithm. Both follow from the solution \eqref{ThetaLogEta} written using spatial infinity slicings. In null coordinates, the pole gives the $1/u$ tail, while the logarithms combine to give the radial logarithm. Using $\eta^2=r^2(1-s^2)$, the $k=1$ contribution is
\begin{multline}
\left.\Phi_{k=1}\right|_{\ell m}
=\frac{Y_{\ell m}}{r^2}\Bigg[
\frac{\Theta_{\ell m}P_\ell(s)-\Lambda_{\ell m}P_{\ell-1}(s)}{1-s^2}\\
+P'_\ell(s)\left(
\frac{\Theta_{\ell m}}{2}\log\frac{1+s}{1-s}
+\Lambda_{\ell m}\log\eta
\right)
+\text{regular terms}
\Bigg]\,.
\end{multline}
The pole and the $Q$-branch logarithm have the following expansions in future and past null coordinates,
\be
\begin{split}
\dI^+_-:&\qquad \frac{1}{1-s^2}=-\frac{r}{2u}+\cO(1)\,,
\qquad \log\frac{1+s}{1-s}=\log\frac{2r}{|u|}+\cO(r^{-1})\,,\\
\dI^-_+:&\qquad\frac{1}{1-s^2}=\frac{r}{2v}+\cO(1)\,,
\qquad \;\;\; \log\frac{1+s}{1-s}=-\log\frac{2r}{v}+\cO(r^{-1})\,.
\end{split}
\ee
At either corner, $\log\eta=\log r+\frac12\log(1-s^2)$, so
$\log\eta$ contributes half a radial logarithm. Evaluating the
Legendre polynomials at the two limits $s\to\pm1$ and summing the
harmonics gives
\begin{equation}
\begin{aligned}
&\chi^+_-=\frac12(\Lambda-\Theta)\,,\qquad &&\Phi^+_-=-\frac14D^2(\Theta+\Lambda)\,,\\
&\chi^-_+(x^A)=\frac12(\Theta+\Lambda)(-x^A)\,,\qquad &&\Phi^-_+(x^A)=\frac14D^2(\Lambda-\Theta)(-x^A)\,.
\end{aligned}
\label{nullDataLogEta}
\end{equation}
The antipodal map at the past corner follows from the parity $P_\ell(-s)=(-1)^\ell P_\ell(s)$. 

Equation \eqref{nullDataLogEta} gives $\frac12D^2\Theta$ in terms of the data at either corner. Denoting the boundary values of $\mathcal Q$ by
\begin{equation}
Q^+[\mu^+]
\equiv
\lim_{s\to+1}\mathcal Q\,,
\qquad
Q^-[\mu^-]
\equiv
\lim_{s\to-1}\mathcal Q\,,
\label{LogCharge}
\end{equation}
where the past charge is written in terms of the antipodally transported function
\begin{equation}
\mu^-(x^A)=\mu^+(-x^A)\,.
\label{muMatching}
\end{equation}
conservation of $\mathcal Q$ gives
\begin{equation}
Q^+[\mu^+]=Q^-[\mu^-]\,,
\label{chargeMatching}
\end{equation}
which is the main equation of this section. This relation will be used at the quantum level to show its equivalence with logarithmic soft scalar emission. To do so, we first establish the boundary conditions for the scattering problem we consider.

\subsubsection*{Scattering states} 

For the classical scattering problem of Section~\ref{sec:first} we impose retarded boundary conditions and no incoming scalar radiation. At $\dI^+$, the retarded roots of all particles are such that at fixed $u$ the field $\Phi$ in the $k=1$ sector admits the radial expansion
\be
\label{eq:fut}
\Phi_{k=1}(u,r,\hat x)
=\frac{\Lambda(\hat x)}{ru}
-\frac{D^2\Lambda(\hat x)}{2r^2}\log|u|
+ O(r^{-2}),
\ee
with $\Lambda$ given by the particular data of the retarded solution. By \eqref{nullDataLogEta} this translates into the condition $\Theta=-\Lambda$ for all modes $\ell\geq1$. However at $\dI^-_+$, the retarded field $\Phi$ changes drastically as it now reads
\be
\label{eq:pas}
\Phi_{k=1}(v,r,\hat x)=\frac{D^2\Lambda(-\hat x)}{2r^2}\log r + O(r^{-2})\,,
\ee
which contains no tail data, only a term $\log r/r^2$ produced by logarithmic corrections to the incoming trajectories in the Coulomb field of the incoming particles. For these states, the charge \eqref{LogCharge}  becomes
\begin{equation}
\cQ^+[\mu^+]
=-\frac12\oint d^2x\,\sqrt{\gamma}\,
\mu^+(x^A)D^2\Lambda(x^A)\,,
\qquad
\cQ^-[\mu^-]
=-\frac12\oint d^2x\,\sqrt{\gamma}\,
\mu^-(x^A)D^2\Lambda(-x^A)\,.
\label{chargesScattering}
\end{equation}
Conservation then requires that the retarded solution satisfies the antipodal
map, which we verify in Appendix~\ref{AppCoulomb}.

In summary, for retarded scattering states, \eqref{chargesScattering}
expresses the same conserved charge in terms of $\Lambda$ on the past and the future. At $\dI^+_-$, $\Lambda$ is the nonconstant part of the $1/u$ tail. At $\dI^-_+$, it is determined
by a subleading radial logarithm at the antipodal point. These two representations provide the starting point for the decomposition into soft and hard contributions.

\subsubsection*{Soft--hard decomposition of the logarithmic charge} 

The soft and hard decomposition separates the flux at null infinity from the terms contributing to the logarithmic mode. We start from \eqref{LogCharge} without imposing the retarded scattering conditions at this stage.
Recalling the definition in \eqref{phi0LargeU}, the tails at future and past null infinity,
respectively, are extracted from the leading radial field as follows
\begin{equation}
\lim_{u\to\pm\infty}
u^2\partial_uD^2\phi_0
=
-D^2\chi^+_\pm\,,
\qquad
\lim_{v\to\pm\infty}
v^2\partial_vD^2\phi_0
=
-D^2\chi^-_\pm\,.
\label{tailExtractionCharge}
\end{equation}
Using \eqref{tailExtractionCharge}, we find
\begin{equation}
Q^\pm[\mu^\pm]
=
Q_S^\pm[\mu^\pm]+Q_H^\pm[\mu^\pm]\,,
\end{equation}
with
\begin{align}
Q_S^+[\mu^+]
&=
-\frac{1}{2}
\int_{\dI^+}
du\,d^2x\,\sqrt{\gamma}\,
\mu^+
\partial_u
\left(u^2\partial_uD^2\phi_0\right),
&
Q_H^+[\mu^+]
&=
-
\int_{\dI_{+}^{+}}d^2x\,\sqrt{\gamma}\,
\mu^+\left(\frac{1}{2}D^2\chi^+_++\Phi^+_-\right)\,,
\label{futureSoftHard}
\\
Q_S^-[\mu^-]
&=
-\frac{1}{2}
\int_{\dI^-}
dv\,d^2x\,\sqrt{\gamma}\,
\mu^-
\partial_v
\left(v^2\partial_vD^2\phi_0\right),
&
Q_H^-[\mu^-]
&=
\int_{\dI_{-}^{-}}d^2x\,\sqrt{\gamma}\,
\mu^-\left(\frac{1}{2}D^2\chi^-_--\Phi^-_+\right).
\label{pastSoftHard}
\end{align}
The soft terms measure the change in the tails along null infinity.
The hard contributions contain the tail $\chi^+_+$ and the logarithmic boundary term $\Phi^+_-$ on the future, and the early-time tail $\chi^-_-$ and the non-radiative logarithm $\Phi^-_+$ on the past side. For the retarded scattering solution, the absence of incoming radiation implies that,
\begin{equation}
Q^-_S[\mu^-]=0\,.
\label{pastCoulombCharge}
\end{equation}
In Sections~\ref {sec:hard} and \ref{sec:Ward}, this condition will be relevant for identifying the precise soft contribution and its action on scattering states.

\subsection{Classical tail memory}
\label{sec:tailmemory}

For the retarded scattering solution of Section~\ref{sec:first}, leading scalar memory compares the early and late values of the
waveform, while tail memory compares the coefficients of its $1/u$ tails. In the absence of incoming radiation, the past charge
is entirely non-radiative. Antipodal matching relates this charge to the early tail at $\dI^+$, so the change in the tail along
$\dI^+$ measures the difference between the two hard charges.

We define the angular tail memory by
\be
\Delta_{\rm tail}(x^A)=
\int_{-\infty}^{+\infty}du\,
\partial_u\!\left[u^2\partial_uD^2\phi_0(u,x^A)\right]\,.
\label{tailMemoryDef}
\ee
The late tail enters the future hard charge through \eqref{futureSoftHard}, and the early tail enters the past hard charge through \eqref{pastSoftHard}. With the antipodally
matched smearings and $Q_S^-=0$, charge conservation gives
\begin{equation}
Q_H^+[\mu^+]-Q_H^-[\mu^-]
=
-Q_S^+[\mu^+]
=
\frac12\int_{S^2}d^2x\,\sqrt{\gamma}\,
\mu^+(x^A)\Delta_{\rm tail}(x^A)\,.
\label{hardChargeMemory}
\end{equation}
For the point-particle trajectories, the tail coefficients in
\eqref{retardedEndpointCoefficients} give
\begin{equation}
\Delta_{\rm tail}^{\rm pp}(x^A)
=
\frac{1}{4\pi}D^2
\left[
\sum_{a\in{\rm out}}
\kappa_a\frac{q\cdot C_{a,+}}{q\cdot V_{a,+}}
-
\sum_{a\in{\rm in}}
\kappa_a\frac{q\cdot C_{a,-}}{q\cdot V_{a,-}}
\right].
\label{tailMemoryC}
\end{equation}

\section{Interacting scalar field theories near timelike infinity}
\label{sec:interacting}

The worldline description identifies the classical content of the tail memory. We now derive the corresponding results in field theory. We characterize, near future timelike infinity $i^{+}$, the effect of the backreaction of the massive sector as a source that couples to the massless scalar. We treat two models: a complex massive scalar with a cubic coupling, and the Dirac fermion of \eqref{YukawaL}. In both cases, the logarithmic part of the source takes the same derivative form once it is expressed through the scalar charge carried by a massive particle.

We use coordinates that parameterize flat space close to $i^{+}$,
\be
x^\mu=\tau Y^\mu(y)\,,\qquad Y^2=-1\,,\quad Y^0>0\,,\qquad
ds^2=-d\tau^2+\tau^2 h_{ab}(y)\,dy^a dy^b\,,
\ee
where $h_{ab}$ is the metric on the unit hyperboloid $\mathbb{H}^3$, with covariant derivative $\nabla_a$, Laplacian $\Delta$ and measure $d^3V=\sqrt h\,d^3y$. For a scalar,
\be
\Box f = -\partial_\tau^2 f - \frac{3}{\tau}\partial_\tau f + \frac{1}{\tau^2}\Delta f\,.
\ee
The long-range field that enters both calculations is the leading term of
the massless scalar near $i^+$,
\be
\Phi(\tau,y)=\frac{\phi_0(y)}{\tau}+\cdots\,.
\label{masslessTimelikeExpansion}
\ee
Here $\phi_0(y)$ denotes the coefficient of $1/\tau$ at timelike infinity, different from the radiative coefficient $\phi_0(u,\hat x)$
at null infinity. We will use the arguments of these functions to distinguish the two uses of the notation. This coefficient $\phi_0(y)$ includes a particular solution proportional to the matter coupling and an independent homogeneous contribution. We
evaluate them separately in Section~\ref{sec:hard}. For the free massive field we start with solutions of the form
\be
\psi(\tau,y)=\frac{e^{\alpha\tau}}{\tau^{3/2}}\sum_{n=0}^\infty\frac{\rho_n(y)}{\tau^n}\,,\qquad \alpha^2=-M^2\,.
\label{MilneExp}
\ee

\subsection{Massive scalar matter}
\label{sec:scalarmatter}

We take a complex massive scalar $\Psi_s$ coupled to $\Phi$ through
\be
\cL_s=-|\d\Psi_s|^2-M^2|\Psi_s|^2-\frac{1}{2}(\d\Phi)^2+g_s\,\Phi\,|\Psi_s|^2\,.
\label{scalarModel}
\ee
The equations of motion are
\be
(\Box-M^2)\Psi_s=-g_s\Phi\Psi_s\,,\qquad
\Box\Phi=-g_s|\Psi_s|^2\,.
\label{scalarMatterEOM}
\ee
The coupling $g_s$ has dimension one. Writing $p_i=MY_i$ in the leading soft factor $g_s/(2p_i\cdot k)$ identifies the scalar charge of each particle as $\kappa=g_s/(2M)$.

We solve the equations perturbatively in the coupling,
\be
\Psi_s=\psi_s+g_s\psi_{s,1}+\cO(g_s^2)\,, \quad
\Phi=\phi+O(g_s)
\ee
leading to
\be
(\Box-M^2)\psi_s=0\,,\qquad
(\Box-M^2)\psi_{s,1}=-\phi\,\psi_s\,.
\ee
Using the long-range field \eqref{masslessTimelikeExpansion}, we allow
logarithms in the correction,
\be
\psi_{s,1}=\frac{e^{\alpha \tau }}{\tau^{3/2}}\sum^{\infty}_{n=0}\frac{b_{n}+ \log(\tau/\tau_0)\,b^L_{n}}{\tau^{n}}\,.
\ee
Matching powers of $\tau^{-1}$ and $\log\tau$ yields the free recursion and the first logarithmic relations,
\be
\rho_1=-\frac{1}{2\alpha}\Big(\Delta+\frac{3}{4}\Big)\rho_0\,,
\qquad
b_0^{L}=\frac{1}{2\alpha}\,\phi_0\,\rho_0\,,
\qquad
b_1^{L}=-\frac{1}{2\alpha}\Big(\Delta+\frac{3}{4}\Big)b_0^{L}\,,
\ee
The logarithmic part of the density correction follows from these coefficients. For the time-independent contribution in \eqref{MilneExp}, the correction to the density $|\Psi_s|^2$ is
\be
\cJ_{\rm corr}=\psi_s^*\psi_{s,1}+\psi_s\psi_{s,1}^*\,,
\qquad
\tau^3\cJ_{\rm corr}=\cJ_0+\frac{\log(\tau/\tau_0)}{\tau}\,\cJ_1+\cO(\tau^{-1})\,.
\label{current}
\ee
Its spatial variation on the hyperboloid gives the first surviving logarithmic
coefficient,
\be
\cJ_{1}=\frac{1}{2M^2}\nabla_a\big(|\rho_0|^2\nabla^a\phi_0\big)\,.
\label{Lscalar}
\ee

\subsection{Yukawa theory}
\label{sec:yukawa}

We now take the Dirac fermion of \eqref{YukawaL}. In coordinates adapted to $i^+$, the Dirac operator is
\be
\slashed{\nabla}=\gamma^0\Big(\partial_\tau+\frac{3}{2\tau}\Big)+\frac{1}{\tau}\gamma^a\nabla_a\,.
\ee
Projecting the free equation with $P_\pm=\frac{1}{2}(1\pm\gamma^0)$ gives
\be
P_+\rho_0=0\,,
\qquad
2M\,P_+\rho_{n+1}=i\gamma^a\nabla_a\rho_n-i n\gamma^0\rho_n\,,
\ee
so that
\be
P_+\rho_1=\frac{i}{2M}\gamma^a\nabla_a\rho_0\,,
\qquad
P_-\rho_1=\frac{i}{2M}\Big(\Delta+\frac{3}{2}\Big)\rho_0\,,
\ee
where we used the spinor identity on the unit hyperboloid, $\gamma^a\gamma^b\nabla_a\nabla_b=-\big(\Delta+\tfrac{3}{2}\big)$. We use the same expansion  \eqref{masslessTimelikeExpansion} to compute the first interaction correction. We write $\Psi=\psi+g\psi_1+\cO(g^2)$ with $(i\slashed\nabla-M)\psi_1=-\phi\,\psi$, and allow
\be
\psi_1=\frac{e^{iM\tau}}{\tau^{3/2}}\sum_{n=0}^\infty\frac{b_n+\log(\tau/\tau_0)\,b_n^L}{\tau^n}\,.
\ee
At order $g$ the corrected spinor $\psi_1$ and the scalar obey
\be
i\gamma^0\!\Big(\partial_\tau+\frac{3}{2\tau}\Big)\psi_1+\frac{i}{\tau}\gamma^a\nabla_a\psi_1-M\psi_1=-\psi\,\phi,
\qquad
\Box\phi=-\cJ_{\rm corr},
\qquad
\cJ_{\rm corr}\equiv \bar{\psi}_1\psi+\bar{\psi}\psi_1.
\ee
Solving order by order in $\tau^{-1}$ one finds the leading relations
\be
P_+ b_0=P_+ b_0^L=0,
\qquad
b_0^L=-i\,\rho_0\,\phi_0,
\qquad
P_+ b_1^L=\frac{1}{2M}\gamma^a\nabla_a(\rho_0\phi_0),
\qquad
P_- b_1^L=\frac{1}{2M}\Big(\Delta+\frac{3}{2}\Big)(\rho_0\phi_0).
\ee
Inserting these into $\cJ_{\rm corr}$ and organizing the result as in \eqref{current}, the logarithmic piece again takes the universal derivative form
\begin{equation}
\cJ_1(Y)=\frac{1}{M}\,\nabla_a\!\Big(|\rho_0|^2\,\nabla^a\phi_0\Big).
\end{equation}

\subsection{Quantum density and the universal logarithmic source}
\label{sec:density}

After including both frequency sectors and normal ordering, the leading massive density of the Yukawa theory is
\be
n(Y)=\frac{M^2}{2(2\pi)^3}\sum_s\Big(b^\dagger_{M\vec Y;s}b_{M\vec Y;s}+d^\dagger_{M\vec Y;s}d_{M\vec Y;s}\Big)\,,
\label{density}
\ee
where the factor $M^2$ includes the spinor normalization $\bar u_su_{s'}=2M\delta_{ss'}$. With
\be
\{b_{\vec p;s},b^\dagger_{\vec p';s'}\}=\{d_{\vec p;s},d^\dagger_{\vec p';s'}\}=2E_{\vec p}(2\pi)^3\delta_{ss'}\delta^{(3)}(\vec p-\vec p')\,,
\ee
its action on a Fock state of massive particles is
\be
n(Y)|X\rangle=\sum_{i\in X}\delta_{\cH}(Y,Y_i)|X\rangle\,,\qquad \int d^3V\,\delta_\cH(Y,Y_i)=1\,.
\label{densityAction}
\ee
The source relevant to the tail is organized as
\be
J(\tau,Y)=\frac{g\,n(Y)}{\tau^3}+\frac{g^2}{\tau^4}\log\frac{\tau}{\tau_0}\,\cJ_1(Y)+\cdots\,,
\label{JMilne}\ee
with the $\cJ_1(Y)$ computed previously and identified as the first interaction correction. The $\tau^{-3}$ term is the universal tree-level source identified in \cite{Campiglia:2017dpg}. The logarithmic current is diagonal in the asymptotic spin through $n(Y)$. 

For massive scalar matter, the normal-ordered sum of the two frequency
densities in \eqref{Lscalar} is $n_s/(2M)$, where the scalar number
density $n_s$ acts as in \eqref{densityAction}. Expressed in terms of the
scalar charge per particle, the logarithmic source has the same form in
both models,
\be
(\text{coupling})^2\times \mathcal{J}_{1}=\frac{\kappa^2}{M}\,\nabla_a\big(n\,\nabla^a\phi_0\big)\,.
\label{universalSource}
\ee
with $\kappa=g$ for Dirac fermions and $\kappa=\tfrac{g_s}{2M}$ for massive complex scalars. Therefore, this derivative structure is the same once each coupling is expressed through the scalar charge carried by a massive particle. In what follows, we continue using only Dirac fermions.

\subsection{Retarded propagation to null infinity}
\label{sec:retarded}

The retarded response to the source \eqref{JMilne} is
\be
\Phi_R(x)=\int_0^\infty d\tau'\,\tau'^3\int d^3V'\,G_R(x-\tau'Y')\,J(\tau',Y')\,,
\qquad
G_R(z)=\frac{1}{2\pi}\theta(z^0)\delta(-z^2)\,.
\label{retardedResponse}
\ee
To obtain the leading term near null infinity, we follow the strategy of \cite{Campiglia:2019wxe}. Using coordinates pointing at $\dI^+$ given by $x=(u+r,r\hat x)$ the leading radial coefficient is therefore
\be
\Phi_R(u,\hat x)=-\frac{1}{4\pi}\int d^3V'\,\frac{\tau'^3J(\tau',Y')}{q\cdot Y'}\bigg|_{\tau'=u/(-q\cdot Y')}\,.
\label{phiRMilne}
\ee
We use this expression for the late-time future region. The coefficient of $1/u$ is encoded in $D^2\chi^+_+$. Using the sphere identity
\be
D^2\log(-q\cdot Y)=\frac{1}{(q\cdot Y)^2}-1\,,
\label{sphereIdentity}
\ee
we obtain
\be
D^2\chi^+_+(\hat x)=\frac{g^2}{4\pi M}\int d^3V\,n(Y)\,\nabla_a\!\left(\frac{1}{(q\cdot Y)^2}\right)\nabla^a\phi_0(Y)
\,.
\label{projectedLateRetardedTail}
\ee
This is the expression directly measured by hard charge $Q_H^+$ as defined in \eqref{futureSoftHard}.

\subsubsection*{Past timelike infinity}

The same analysis applies near $i^-$, where we write $x=-\tau Y$ with $Y$ future-directed, so an incoming particle of momentum $p_i$ sits at $Y_i=p_i/M$. The asymptotic equations and the source take the same form, and the coefficient of $1/u$ at $u\to-\infty$ satisfies
\be
D^2\chi^+_-(\hat x)=-\frac{g^2}{4\pi M}\int d^3V\,n_{\rm in}(Y)\,\nabla_a\!\left(\frac{1}{(q\cdot Y)^2}\right)\nabla^a\tilde\phi_0(Y)\,,
\label{earlyRetardedTail}
\ee
where $n_{\rm in}$ is the density of incoming particles and $\tilde\phi_0$ the coefficient of the leading $1/\tau$ scalar field near past-like infinity $i^-$.

\section{The hard charge}
\label{sec:hard}

We now compute how the hard charges act on external states.

\subsection{Action on massive states}

We start with the charge defined at future. Inserting the tail
\eqref{projectedLateRetardedTail} into the value of the hard charge in the future \eqref{futureSoftHard} and using the action of the number operators on hard states 
\eqref{densityAction}, we obtain
\be
\begin{aligned}
\langle{\rm out}|Q_H^+[\mu^+]\,S|{\rm in}\rangle
={}&-\frac{g^2}{8\pi M}\int_{S^2}d^2x\,\sqrt{\gamma}\,\mu^+
D^2\sum_{i\in{\rm out}}
\langle{\rm out}|\cD_i(q)\phi_0(Y_i)\,S|{\rm in}\rangle\,.
\end{aligned}
\label{QHoutStates}
\ee
Here we used $\Phi^+_-=0$ for the states considered, as shown in Appendix~\ref{AppCoulomb}. The differential operator $\cD_i(q)$ on the hyperboloid is the one introduced in \eqref{Tdef} for the particle calculation. The sphere Laplacian removes the constant mode of the tail. It is important to notice that the expression above is normal ordered and the scattering prescription fixes the ordering of the density and field.

The contributions to the hard charge follow from the leading equation for
the scalar at timelike infinity. Using the source in \eqref{JMilne},
we can express the solution as
\be
(\Delta+1)\phi_0=-g\,n\,,\qquad
\phi_0=\phi_0^{\rm P}+\phi_0^{\rm H}\,,\qquad
(\Delta+1)\phi_0^{\rm H}=0\,\,
\label{scalarTimelike}
\ee
where $\rm P$ and $\rm H$ stand for particular and homogeneous solutions.
\subsection{Particular contribution}
\label{sec:classical}

To compare with the classical trajectories, we evaluate the
particular solution on each outgoing particle. The hyperbolic
Green function for the leading source is \cite{Campiglia:2017dpg}
\be
\phi_0^{\rm P}(Y)=\frac{g}{4\pi}\int d^3V'\,
\frac{n(Y')}{\sqrt{(Y\cdot Y')^2-1}}\,.
\label{phiP}
\ee
On an outgoing state, this gives the field sourced by the other
outgoing particles. Evaluated on particle $i$, it reads
\begin{align}
\phi_0^{\rm P}(Y_i)
&=\frac{g}{4\pi}\sum_{\substack{j\in{\rm out}\\ j\ne i}}\frac{1}{\sqrt{(Y_i\cdot Y_j)^2-1}}, 
\end{align}
Since this field is diagonal on $\langle \mathrm{out}|$, inserting it into \eqref{QHoutStates} gives
\begin{align}
    -\frac{g^3}{32\pi^2 M}\int_{S^2}d^2x\,\sqrt{\gamma}\,\mu^+ D^2\sum_{\substack{i,j\in{\rm out}\\ i\ne j}}
 \cD_i(q) \left[\frac{1}{\sqrt{(Y_i\cdot Y_j)^2-1}} \right]\langle{\rm out}|S|{\rm in}\rangle\,,
\end{align}
This is the same sum over pairs as in the tail trajectory calculation
\eqref{projectedParticleTail}, including the mass and coupling factors,
\be
    D^2\chi^{+,\rm P}_+
=\frac{g^3}{16\pi^2 M}\,D^2
\sum_{\substack{i,j\in{\rm out}\\ i\ne j}} \cD_i(q) \left[\frac{1}{\sqrt{(Y_i\cdot Y_j)^2-1}} \right]
\label{classicalMatch}
\ee
Thus the particular solution accounts for the classical hard charge.
This has a counterpart in the amplitude calculation, where replacing
the Feynman propagator by the retarded one isolates the classical
logarithm \cite{Sahoo:2018lxl}\footnote{See \cite{AtulBhatkar:2021txo}, for a discussion on the use of the Feynman propagator in QED and its relation with asymptotic charges.}. The difference between these propagators
solves the homogeneous wave equation. We next evaluate the homogeneous
field insertion between scattering states.

\subsection{Homogeneous contribution}
\label{sec:quantum}
The homogeneous contribution comes from the free part of $\phi$ near future timelike infinity, where we use the coordinate chart that parameterizes $i^+$. In terms of outgoing oscillators,
\be
\phi^{\rm H}(\tau Y)=\int\frac{d^2\hat x'\,\sqrt{\gamma}}{2(2\pi)^3}\int_0^\infty d\omega\,\omega\,
\Big[a_{\rm out}(\omega\hat x')\,e^{i\tau\omega\,Y\cdot q'}+a^\dagger_{\rm out}(\omega\hat x')\,e^{-i\tau\omega\,Y\cdot q'}\Big]\,.
\ee
Since $Y$ is timelike and $q'$ is null, $Y\cdot q'<0$ over the whole sphere. The frequency integral therefore has no stationary point, and at large $\tau$ it is dominated by its lower endpoint. Assuming that $\omega\,a_{\rm out}(\omega\hat x')$ has a finite limit as $\omega\to0$, this gives an expansion of the form \eqref{masslessTimelikeExpansion}, where the $1/\tau$ coefficient of the homogeneus field is
\be
\phi^{\rm H}_0(Y)=\,\lim_{\omega\to0^+} \frac{i}{16\pi^3}\oint d^2\hat x'\,\frac{\sqrt{\gamma}}{Y\cdot q'}\omega\big(a_{\rm out}(\omega\hat x')-a^\dagger_{\rm out}(\omega\hat x')\big)\,.
\label{psiquantum}
\ee
This mode is the zero-frequency mode of the outgoing free field, which is also controlled by the leading soft theorem. On the outgoing states, $a^\dagger_{\rm out}$ annihilates the state, while $a_{\rm out}(\omega\hat x')$ adds an outgoing scalar of momentum $\omega q'$. The leading soft relation \eqref{leadingSoft} then gives
\be
\langle{\rm out}|\phi_0^{\rm H}(Y_i)\,S|{\rm in}\rangle
=\frac{igM^2}{8\pi^2}\sum_j\eta_j\,I^{\rm q}_{ij}\,\cA_N\,.
\label{phiHmatrix}
\ee
after inserting \eqref{psiquantum} at $Y_i=p_i/M$ with $\cA_N=\langle{\rm out}|S|{\rm in}\rangle$. The two-particle function $I^{\rm q}_{ij}$ is the angular integral of the two soft
denominators. The evaluation 
\cite{Weinberg:1995mt} gives
\be
\begin{split}
I^{\rm q}_{ij}
&=\frac{1}{2\pi}\int_{S^2}d^2x'\,\sqrt{\gamma}
\frac{1}{(p_i\cdot q')(p_j\cdot q')}\\
&=\frac{1}{\sqrt{(p_i\cdot p_j)^2-M^4}}
\log\!\left[
\frac{-p_i\cdot p_j+\sqrt{(p_i\cdot p_j)^2-M^4}}
     {-p_i\cdot p_j-\sqrt{(p_i\cdot p_j)^2-M^4}}
\right].
\end{split}
\label{Iq}
\ee
This insertion affects all the hard particles because the soft scalar
can attach to an incoming or an outgoing leg. 

\subsection{The past hard charge}
\label{sec:pasthard}

For the in-states considered here, the past hard charge is read from the non-radiative data, as in \eqref{pastCoulombCharge}. We use
antipodal matching to express it on the future sphere, where
\eqref{earlyRetardedTail} gives its action in terms of the incoming
massive density. With the smearings matched as in \eqref{muMatching},
\be
\begin{aligned}
\langle{\rm out}|S\,Q_H^-[\mu^-]|{\rm in}\rangle
={}&+\frac{g^2}{8\pi M}\int_{S^2}d^2x\,\sqrt{\gamma}\,\mu^+
D^2\sum_{i\in{\rm in}}
\langle{\rm out}|S\,\cD_i(q)\tilde\phi_0(Y_i)|{\rm in}\rangle\,,
\end{aligned}
\label{QHinStates}
\ee
where $\tilde\phi_0$ is the field evaluated in the past timelike infinity. The sign follows from the early-time tail. For the particular
solution we use the same Green function as in \eqref{phiP}, now
sourced by incoming particles,
\be
\tilde\phi_0^{\rm P}(Y_i)=\frac{g}{4\pi}
\sum_{\substack{j\in{\rm in}\\j\ne i}} \frac{1}{\sqrt{(Y_i\cdot Y_j)^2-1}}\,.
\ee
This reproduces the early-time tail of the incoming
trajectories, by the same comparison as in \eqref{classicalMatch}.

The past homogeneous field is obtained in the same way from the free incoming field near past timelike infinity, $x^\mu=-\tau Y^\mu$ with $\tau\to+\infty$. The phase of the frequency integral is reversed, and
\be
\tilde\phi^{\rm H}_0(Y)=-\lim_{\omega\to0^+} \frac{i}{16\pi^3}\oint d^2\hat x'\,\frac{\sqrt\gamma}{Y\cdot q'}\,\tilde \omega\big(a_{\rm in}(\omega\hat x')-a^\dagger_{\rm in}(\omega\hat x')\big) \,,\,.
\ee
On a ket with no incoming scalars, $a_{\rm in}$ annihilates the state, while $a^\dagger_{\rm in}(\omega\hat x')$ adds an incoming scalar of momentum $\omega q'$. This analysis yields
\be
\langle{\rm out}|S\,\tilde\phi_0^{\rm H}(Y_i)|{\rm in}\rangle
=-\frac{igM^2}{8\pi^2}\sum_j\eta_j\,I^{\rm q}_{ij}\,\cA_N\,.
\label{phiHmatrixPast}
\ee

\subsection{Combined kernel}

To compare the charges with amplitudes, we express the derivative along the hyperboloid in terms of the on-shell momenta. The completeness relation in \eqref{Tdef} gives the orbital angular momentum generator
contracted with the soft direction,
\be
\cD_i(q)f(MY_i)=\frac{q^\mu p_i^\nu}{q\cdot p_i}J^{(i)}_{\mu\nu}f(p_i)\,,\quad
J^{(i)}_{\mu\nu}=p_{i\mu}\frac{\d}{\d p_i^\nu}
                 -p_{i\nu}\frac{\d}{\d p_i^\mu}\,.
\label{Ddef}
\ee
In the same variables, the classical kernel is
\be
I^{\rm cl}_{ij}=\frac{1}{\sqrt{(p_i\cdot p_j)^2-M^4}}\,.
\label{Icl}
\ee
The particular solution couples particles on the same side of the
scattering process. The homogeneous field couples to both sides, with
the signs fixed by emission and absorption. For distinct legs, the two contributions combine into\footnote{The classical self-term is subtracted in the particular retarded solution. In
the homogeneous field the coincident function $I^{\rm q}_{ii}$ is constant on the mass shell, so its orbital derivative vanishes.}
\be
I^{(\eta_i)}_{ij}=\delta_{\eta_i,\eta_j}\,I^{\rm cl}_{ij}
+\frac{i\eta_i\eta_j}{2\pi}\,I^{\rm q}_{ij}\,.
\label{Icombined}
\ee
Substituting the two fields into \eqref{QHoutStates} and \eqref{QHinStates}, we obtain
\begin{align}
\langle{\rm out}|Q_H^+[\mu^+]\,S|{\rm in}\rangle
&=-\frac{g^3M}{32\pi^2}
\int_{S^2}d^2x\,\sqrt{\gamma}\,\mu^+D^2
\sum_{i\in{\rm out}}\sum_{j\ne i}
\left[\frac{q^\mu p_i^\nu}{q\cdot p_i}J^{(i)}_{\mu\nu}I^{(\eta_i)}_{ij}\right]\,\cA_N\,,
\label{QHoutKernel}\\
\langle{\rm out}|S\,Q_H^-[\mu^-]|{\rm in}\rangle
&=+\frac{g^3M}{32\pi^2}
\int_{S^2}d^2x\,\sqrt{\gamma}\,\mu^+D^2
\sum_{i\in{\rm in}}\sum_{j\ne i}
\left[\frac{q^\mu p_i^\nu}{q\cdot p_i}J^{(i)}_{\mu\nu}I^{(\eta_i)}_{ij} \right]\,\cA_N\,.
\label{QHinKernel}
\end{align}
These expressions give the hard side of the Ward relation. In Section~\ref{sec:oneloop} we compute the logarithmic soft emission directly from the one-loop amplitude and check that its projection on the sphere gives the same expressions as the ones obtained here.

\section{Scalar soft emission at one loop}
\label{sec:oneloop}

The future and past hard charges in Section~\ref{sec:hard} are expressed using an orbital derivative acting on a common contribution that depends on a pair of external particles. In this section, we compute the logarithmic contribution to the one-loop soft emission directly in the Yukawa theory and compare with the result obtained using the action of the hard charge derived before.

We denote the soft emission of a scalar particle with momentum $k$. We split the computation into diagrams producing dipole interactions and single-particle contributions.

\subsection{Dipolar diagrams}
\label{sec:dipolar_diagrams}

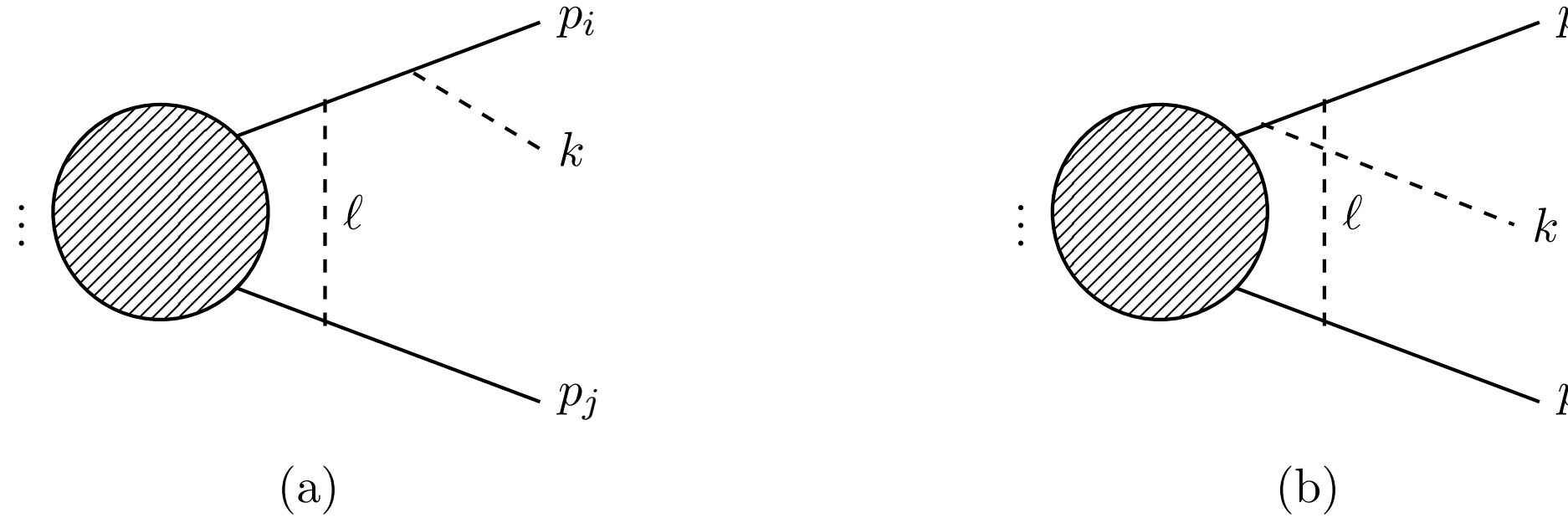
\begin{figure}[h!]
\centering
\begin{minipage}{0.45\textwidth}
\centering
\begin{tikzpicture}
\draw[thick] (0.6,0.6) -- (3,1.5)
    node[pos=1, right] {$p_i$};
\draw[thick] (0.6,-0.6) -- (3,-1.5)
    node[pos=1, right] {$p_j$};
\draw[dashed,thick] (1.3,-0.9) -- (1.3,0.9)
    node[midway, right] {$\ell$};
\draw[dashed,thick] (2,1.1) -- (3,0.5)
    node[pos=1, right] {$k$};
\node[
draw,
circle,
pattern=north east lines,
thick,
minimum size=1.7cm
] at (0,0) {};
\node at (-1.1,0) {$\vdots$};
\end{tikzpicture}

(a)
\end{minipage}
\begin{minipage}{0.45\textwidth}
\centering
\begin{tikzpicture}
\draw[thick] (0.6,0.6) -- (3,1.5)
    node[pos=1, right] {$p_i$};
\draw[thick] (0.6,-0.6) -- (3,-1.5)
    node[pos=1, right] {$p_j$};
\draw[dashed,thick] (1.3,-0.9) -- (1.3,0.9)
    node[midway, right] {$\ell$};
\draw[dashed,thick] (0.8,0.7) -- (2.8,-0.1)
    node[pos=1, right] {$k$};
\node[
draw,
circle,
pattern=north east lines,
thick,
minimum size=1.7cm
] at (0,0) {};
\node at (-1.1,0) {$\vdots$};
\end{tikzpicture}

(b)
\end{minipage}
\caption{Two orderings of soft scalar emission from line $i$ in the presence of virtual scalar exchange with line $j$. The blob denotes the hard amplitude.}
\label{fig:pairwise}
\end{figure}

Contributions containing two external particles interacting at one loop arise from virtual scalar exchange between the two hard lines. The two orderings of the soft emission are displayed in Figure~\ref{fig:pairwise}.

For simplicity, we assume that the $j$-line is an incoming fermion, while the $i$-line corresponds to an outgoing fermion. The momentum on line $i$ is $p_i+k$ and then $p_i+k+\ell$ in diagram (a), and $p_i+\ell$ and then $p_i+k+\ell$ in diagram (b), reading from the external leg towards the hard amplitude. Line $j$ carries $p_j+\ell$. With $p_i^2=-M^2$ and $k^2=0$, the denominators on line $i$ are
\be
\begin{aligned}
(p_i+k)^2+M^2&=2p_i\cdot k\,,\\
\Delta_i(\ell)&=2p_i\cdot\ell+\ell^2\,,\\
\Delta_{ik}(\ell)&=\Delta_i(\ell)+2p_i\cdot k+2k\cdot\ell\,,
\end{aligned}
\ee
and the numerators are
\be
\begin{aligned}
N_a&=\bar u(p_i)(-\slashed p_i-\slashed k+M)
                     (-\slashed p_i-\slashed k-\slashed\ell+M)\,,\\
N_b&=\bar u(p_i)(-\slashed p_i-\slashed\ell+M)
                     (-\slashed p_i-\slashed\ell-\slashed k+M)\,.
\end{aligned}
\ee
The two diagrams read, schematically,
\be
\begin{aligned}
\cA_{(a)}&=g^3\int\frac{d^4\ell}{(2\pi)^4}
\frac{N_a\,H\,\cN_j}
{\ell^2(2p_i\cdot k)\Delta_{ik}(\ell)\Delta_j(\ell)}\,,\\
\cA_{(b)}&=g^3\int\frac{d^4\ell}{(2\pi)^4}
\frac{N_b\,H\,\cN_j}
{\ell^2\Delta_i(\ell)\Delta_{ik}(\ell)\Delta_j(\ell)}\,.
\end{aligned}
\label{diagrams}
\ee
Here $\Delta_j(\ell)=2p_j\cdot\ell+\ell^2$ and $\cN_j=(2M-\slashed\ell)\,u(p_j)$. The hard amplitude $H$ is common to the two orderings, with momenta $p_i+k+\ell$ and
$p_j+\ell$ entering it on the two lines. 

\subsubsection*{Integral evaluation}

We now evaluate the sum of the two diagrams and isolate the terms that will be compared with the action of the hard charge. To do so, we use dimensional regularization with $d=4-2\varepsilon$ with $\varepsilon<0$ to regulate IR divergences of the loop integrals. The coefficient of $\log(\omega)$ is extracted after expanding in $\varepsilon$. 

Since we are interested in the low momentum region of the integral for soft emissions,
we can approximate $H(p_i+k+\ell,p_j+\ell)\approx H(p_i,p_j)$, so that this function is taken out of the integral. After this replacement, the sum of the two diagrams reduces to a set of six integrals that are contracted with the tensors depending on the external momenta and the associated spinor structures. The details of this construction are presented in the Appendix \ref{matias}.  As we want to compare with the analysis of the previous section, we extract terms proportional to $\log(\omega)$ with the presence of the pole $\tfrac{1}{p_i\cdot q}$, where $\omega$ and $q(\hat{x})$  are defined through \eqref{kinematics}. This analysis yields
\be
D^2 L_\omega [\mathcal{A}_{(a)}+\mathcal{A}_{(b)}] =-\frac{i g^3 M^3}{8\pi^2}\,D^2\left[\frac{1}{p_i\cdot k}\left(k\cdot\frac{\partial}{\partial p_i}\right) I^{\rm q}_{ij}\right]\;\bar{u}(p_i)\,H\,u(p_j)\,.
\ee
Using $p_i^2=-M^2$ and \eqref{Ddef}, we write the derivative in terms of the orbital angular momentum as
\be
\frac{M^2\,k\cdot\d_{p_i}}{p_i\cdot k}
=\frac{q^\mu p_i^\nu}{q\cdot p_i}J^{(i)}_{\mu\nu}-p_i\cdot\d_{p_i}\,.
\ee
The radial derivative $p_i\cdot\d_{p_i}$ is independent of $\hat x$ and is removed by
$D^2$. 

The same computation applies to every other pair of external particles, with the momentum adapted to incoming or outgoing lines\footnote{When $i$ and $j$ lie on the same
side of the process, the loop integral develops an
imaginary part. This is precisely what generates the
classical term $\delta_{\eta_i,\eta_j}I^{\rm cl}_{ij}$
in \eqref{Icombined}.}.  
Therefore,
summing over all pairs and using \eqref{Lomega}, we find 
\be
D^2L_\omega\big[\cA_{N+1}\big]
=-\frac{ig^3M}{4\pi}\,
D^2\sum_{i}\sum_{j\ne i}
\left[\frac{q^\mu p_i^\nu}{q\cdot p_i}J^{(i)}_{\mu\nu}I^{(\eta_i)}_{ij}\right]\,\cA_N\,,
\label{DRden}
\ee
where
$\mathcal{A}_N\equiv i \bar{u}(p_i)Hu(p_j)$ and 
$\cA_{N+1}$ stands for the amplitude with the emission of a soft scalar in the presence of $N$ hard fermions, and $I^{(\eta_i)}_{ij}$ is defined in \eqref{Icombined}. The terms in the squared brackets are those of the
future and past hard charges in \eqref{QHoutKernel} and \eqref{QHinKernel}.

\subsection{Single-line terms}

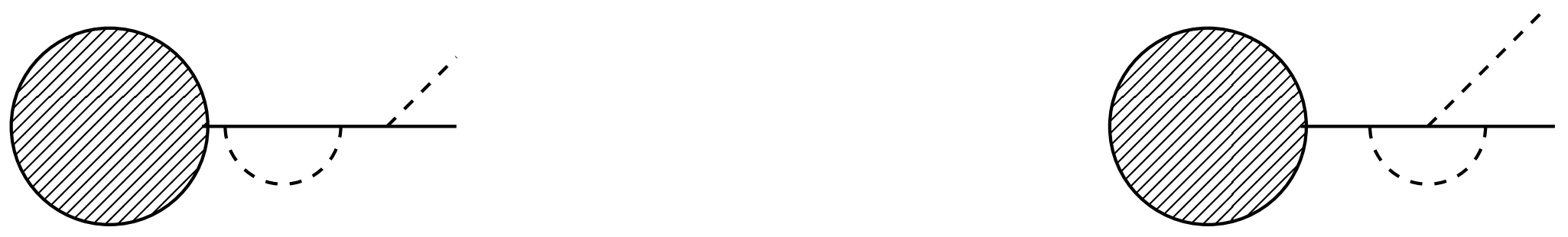
\begin{figure}[htbp]
\centering
\begin{minipage}[t]{0.45\textwidth}
\centering
\begin{tikzpicture}
\draw[thick] (0.8,0) -- (3,0);
\draw[dashed,thick] (2.4,0) -- (3,0.6);
\draw[dashed,thick] (1,0)
    arc[start angle=180,end angle=360,radius=0.5];
\node[
    draw,
    circle,
    pattern=north east lines,
    thick,
    minimum size=1.7cm
] at (0,0) {};
\end{tikzpicture}

\small (a) Self-energy insertion
\end{minipage}\hfill
\begin{minipage}[t]{0.45\textwidth}
\centering
\begin{tikzpicture}
\draw[thick] (0.8,0) -- (3,0);
\draw[dashed,thick] (1.9,0) -- (2.9,1);
\draw[dashed,thick] (1.4,0)
    arc[start angle=180,end angle=360,radius=0.5];
\node[
    draw,
    circle,
    pattern=north east lines,
    thick,
    minimum size=1.7cm
] at (0,0) {};
\end{tikzpicture}

\small (b) Yukawa vertex correction
\end{minipage}
\caption{One-loop corrections on the emitting fermion line. The dashed
arc represents the virtual scalar, and the dashed external line
represents the emitted scalar of momentum $k$.}
\label{fig:singleline}
\end{figure}

Figure~\ref{fig:singleline} shows the two one-loop corrections to soft
scalar emission from a single fermion line. The self-energy dresses
the propagator of momentum $p=p_i+k$, while the vertex correction
joins the fermion line on either side of the emission vertex.

Appendix~\ref{AppRen} evaluates both insertions. The logarithmic terms,
including those proportional to $\slashed{k}$, cancel in their sum
independently of the finite part of the vertex counterterm. That finite
part can then be chosen to express the leading soft pole in terms of
the renormalized coupling.

\section{Scalar Ward identity}
\label{sec:Ward}

We now express the conservation law \eqref{chargeMatching} as a Ward identity for scattering amplitudes and verify it at one loop.
The logarithmic soft charge inserts the coefficient of $\log\omega$ in the emission amplitude. The hard charges give the same angular quantity in terms of dipolar contributions evaluated on the incoming and outgoing hard momenta.

\subsection{Soft charges in terms of oscillators}

To identify the soft insertion, we write the future radiative coefficient in terms of outgoing oscillators,
\be
\phi_0(u,\hat x)
=\frac{1}{8\pi^2i}\int_0^\infty d\omega\,
\Big[
a_{\rm out}(\omega\hat x)e^{-i\omega u}
-a^\dagger_{\rm out}(\omega\hat x)e^{i\omega u}
\Big]\,.
\label{qq}
\ee
After separating any prescribed leading logarithmic branch, the operator $L_\omega$ extracts the coefficient associated with the tails,
\be
a^{\rm out}_{\log}(\hat x)
=L_\omega\big[a_{\rm out}(\omega\hat x)\big]
=\lim_{\omega\to0}\omega\,\d_\omega^2
\big[\omega\,a_{\rm out}(\omega\hat x)\big]\,.
\ee
Following \cite{Campiglia:2019wxe,AtulBhatkar:2019vcb},
we adopt the positive-frequency prescription for inserting soft charge, taking $\omega\to0^+$. This choice is consistent with the retarded propagation used in our derivation and selects the outgoing scalar emission amplitude. The soft insertion becomes
\be
Q^+_S[\mu^+]
=\frac{i}{8\pi}\int_{S^2}d^2x\,\sqrt{\gamma}\,
\mu^+D^2a^{\rm out}_{\log}\,.
\label{softcharge}
\ee
\noindent At past null infinity, the stationary-phase expansion in the
advanced coordinates gives the opposite overall sign and evaluates the
incoming oscillator at the antipodal momentum direction,
\be
\tilde\phi_0(v,\hat x)
=-\frac{1}{8\pi^2i}\int_0^\infty d\omega\,
\Big[
a_{\rm in}(-\omega\hat x)e^{-i\omega v}
-a^\dagger_{\rm in}(-\omega\hat x)e^{i\omega v}
\Big]\,.
\ee
With $a^{\rm in}_{\log}(\hat x)
=L_\omega[a_{\rm in}(\omega\hat x)]$, the past soft charge is
\be
Q^-_S[\mu^-]
=-\frac{i}{8\pi}\int_{S^2}d^2x\,\sqrt{\gamma}\,
\mu^-D^2a^{\rm in}_{\log}(-\hat x)\,.
\label{softchargePast}
\ee
For the scattering states considered here, the incoming state has
no incoming scalar radiation, so we impose
\be
\label{noin}
Q_S^-|{\rm in}\rangle=0\,.
\ee

\subsection{The logarithmic Ward identity}

The charge matching condition \eqref{chargeMatching} gives the Ward identity
\be
\langle{\rm out}|Q^+_S[\mu^+]\,S|{\rm in}\rangle
= -\langle{\rm out}|Q_H^+[\mu^+]\,S|{\rm in}\rangle
+\langle{\rm out}|S\,Q_H^-[\mu^-]|{\rm in}\rangle \,,
\qquad
\mu^-(\hat x)=\mu^+(-\hat x)\,,
\label{Ward}
\ee
where we have used the no-incoming radiation condition. On the outgoing bra, $a_{\rm out}$ adds the soft scalar, so the soft side of
\eqref{Ward} is the logarithmic coefficient of the emission amplitude,
\be
\langle{\rm out}|Q_S^+[\mu^+]\,S|{\rm in}\rangle
=\frac{i}{8\pi}\int_{S^2}d^2x\,\sqrt{\gamma}\,
\mu^+D^2L_\omega\big[\cA_{N+1}(\omega\hat x)\big]\,.
\label{A}
\ee
The hard matrix elements in \eqref{QHoutKernel} and
\eqref{QHinKernel} are already written on the future sphere with
the same smearing $\mu^+$. Inserting \eqref{QHoutKernel} and \eqref{QHinKernel} into the right-hand side of the Ward identity gives
\be
-\langle{\rm out}|Q_H^+[\mu^+]\,S|{\rm in}\rangle
+\langle{\rm out}|S\,Q_H^-[\mu^-]|{\rm in}\rangle
=\frac{g^3M}{32\pi^2}\int_{S^2}d^2x\,\sqrt{\gamma}\,
\mu^+D^2
\sum_i\sum_{j\ne i}
\left[\frac{q^\mu p_i^\nu}{q\cdot p_i}J^{(i)}_{\mu\nu}I^{(\eta_i)}_{ij}\right]\,\cA_N\,.
\label{B}
\ee
By equating \eqref{A} and \eqref{B}, we show that the Ward identity recovers the logarithmic soft one-loop factorization \eqref{DRden} at order $g^3$ for the projected logarithmic coefficient, for all antipodally related smearing functions $\mu$.

\section{Discussion}
\label{sec:discussion}

We have constructed a logarithmic scalar charge from the first subleading $Q$-branch at spatial infinity. Its conservation follows from the asymptotic field equation, and its boundary values give an antipodally matched pair of charges at null infinity. The field equations relate this charge to the sphere Laplacian of the $1/u$ tail, which leads to a soft--hard decomposition and to a corresponding combination of tail-memory data. The logarithmic soft factorization, reproduced by the one-loop calculation, can be confirmed by using the conservation law \eqref{chargeMatching} as a Ward identity. 

The scalar example is useful precisely because the construction does not start from an improper gauge transformation. Similar results in gauge theories rely heavily on the symmetry; here the conservation law follows from asymptotic dynamics and matching, with the masslessness of the field imposed through renormalization conditions. Identifying a transformation generated by the charge would require treating the phase space with logarithmic boundary conditions, possibly in a dual formulation \cite{Henneaux:2018mgn,Campiglia:2018see,Romoli:2024hlc,Romoli:2025map} or by analyzing the canonical formulation of the subleading components that approach $i^0$.  

The analysis presented here can be extended in several directions. The higher logarithmic coefficients suggest further projected conservation laws. Additional spatial logarithms may enlarge the allowed boundary data and alter their matching. A systematic classical limit calculation of the scattering observables, for example through the formalism of \cite{Kosower:2018adc,Manu:2020zxl}, could provide another way to organize the worldline presented in Section~\ref{sec:first} and its relation to the charge conservation correspondence obtained in Section~\ref{sec:charges}. 
Finally, it would be interesting to apply the same strategy to soft relations that are not obviously tied to an asymptotic symmetry, such as one-loop soft theorems in Yang--Mills theory, where collinear singularities must also be taken into account.

\vspace{.5cm}
\paragraph{Acknowledgements}
~\\[8pt]
It is a pleasure to thank Miguel Campiglia, Felipe Díaz, Oscar Fuentealba, Marc Henneaux, Alfredo Pérez, Francisco Rojas, Biswajit Sahoo and Jakob Salzer for useful discussions. This work is funded by the ANID FONDECYT grant 1230853.

\appendix

\section{One loop integrals}
\label{matias}

In terms of the definitions $\Delta_i(\ell)=2p_i\cdot\ell+\ell^2$ and
$\Delta_{ik}(\ell)=\Delta_i(\ell)+2p_i\cdot k+2k\cdot\ell$, the sum of both diagrams can be expressed as
\begin{equation}
\begin{aligned}
\mathcal A_a+\mathcal A_b
&\sim
\int\frac{d^4\ell}{(2\pi)^4}\,
\frac{\bar u(p_i)}{\ell^2\Delta_j(\ell)}
\\
&\quad\times
\Biggl[
\frac{2M}{2p_i\cdot k}\,
\frac{2M-\slashed{k}-\slashed{\ell}}{\Delta_i(\ell)}
\left(1-\frac{2k\cdot\ell}{\Delta_{ik}(\ell)}\right)
-\frac{2}{\Delta_{ik}(\ell)}
\\
&\qquad\quad
+\frac{\slashed{k}\slashed{\ell}}
{(2p_i\cdot k)\Delta_{ik}(\ell)}
+\frac{\slashed{\ell}\slashed{k}}
{\Delta_i(\ell)\Delta_{ik}(\ell)}
\Biggr]
H(2M-\slashed{\ell})u(p_j).
\end{aligned}
\label{fullintegral}
\end{equation}
with $H=H(p_i+k+\ell,p_j+\ell)$ being the hard part that is contained in the blob.
We can reorganize the sum of the two diagrams by noticing that the full result can be written in terms of 6 integrals
\begin{equation}
\begin{alignedat}{2}
I_1&=\int\frac{d^d\ell}{(2\pi)^d}\,\frac{1}{\ell^2\,\Delta_j(\ell)\,\Delta_{ik}(\ell)}\,,
&\qquad
K_0&=\int\frac{d^d\ell}{(2\pi)^d}\,\frac{1}{\ell^2\,\Delta_j(\ell)\,\Delta_i(\ell)}\,,\\[4pt]
I_2^\mu&=\int\frac{d^d\ell}{(2\pi)^d}\,\frac{\ell^\mu}{\ell^2\,\Delta_j(\ell)\,\Delta_{ik}(\ell)}\,,
&
J_2^\mu&=\int\frac{d^d\ell}{(2\pi)^d}\,\frac{\ell^\mu}{\ell^2\,\Delta_j(\ell)\,\Delta_{ik}(\ell)\,\Delta_i(\ell)}\,,\\[4pt]
I_3^{\mu\nu}&=\int\frac{d^d\ell}{(2\pi)^d}\,\frac{\ell^\mu\ell^\nu}{\ell^2\,\Delta_j(\ell)\,\Delta_{ik}(\ell)}\,,
&
J_3^{\mu\nu}&=\int\frac{d^d\ell}{(2\pi)^d}\,\frac{\ell^\mu\ell^\nu}{\ell^2\,\Delta_j(\ell)\,\Delta_{ik}(\ell)\,\Delta_i(\ell)}\,.
\end{alignedat}
\label{loopIntegrals}
\end{equation}
Notice that the $\log(\omega)$ can only come from the evaluation of integrals, and also notice that all terms next to $K_0$ will not generate these logarithmic terms, as that integral does not depend on $k$. In the terms of \eqref{fullintegral} where $2k\cdot\ell$ multiplies additional factors of $\slashed\ell$, we use
\be
\frac{2k\cdot\ell}{\Delta_i\Delta_{ik}}
=\frac{1}{\Delta_i}-\frac{1}{\Delta_{ik}}-\frac{2p_i\cdot k}{\Delta_i\Delta_{ik}}\,,
\label{integralExpansion}
\ee
while the term linear in $\ell$ is kept in its original form. Collecting the numerators by powers of the loop momentum, the integral \eqref{fullintegral} can be written as
\begin{align}
 \mathcal{A}_a+\mathcal{A}_b= g^3 \bar{u}(p_i)&\left\{ -4M H I_1 + \left(- \frac{2M}{2p_i\cdot k}(2M-\slashed{k})\{H,\gamma_\mu\}+2H \gamma_\mu \right)I_2^\mu  \right.\nonumber\\ 
    &+\frac{1}{2p_i\cdot k }(2M-\slashed{k})\gamma_\mu H \gamma_\nu I_3^{\mu \nu} + \left( -2M\{(2M-\slashed{k})H,\gamma_\mu\} - \frac{8M^2}{2p_i\cdot k}(2M-\slashed{k})H k_{\mu} \right)J_2^\mu \nonumber\\
    &+ \gamma_\mu \left(2M-\slashed{k} \right)H \gamma_\nu J_3^{\mu \nu} \Bigg. \Bigg\} u(p_j)+{\rm terms}\,{\rm proportional}\,{\rm to}\,K_0.
\end{align}
In principle, $H$ depends on the loop momentum and remains inside the integral, but since the logarithm of $\omega$ arises from the region of small loop momentum, we may replace $H$ by its value at $k=\ell=0$, $H\to H(p_i,p_j)$\footnote{It is also assumed that $H$ contains the pure hard part.}.

The dependence on $\omega$ is then carried by the integrals \eqref{loopIntegrals} through $\Delta_{ik}$. As $\omega\to0$, each of them contains a term proportional to $\log(\omega/M)$, with a coefficient that scales with $\omega$ as
\begin{equation}
\begin{alignedat}{2}
I_1&=\bar I_1\,\log\frac{\omega}{M}+\dots\,,
&\qquad
J_2^\mu&=\bar J_2^\mu\,\log\frac{\omega}{M}+\dots\,,\\
I_2^\mu&=(p_i\cdot q)\,\bar I_2^\mu\,\omega\,\log\frac{\omega}{M}+\dots\,,
&
J_3^{\mu\nu}&=(p_i\cdot q)\,\bar J_3^{\mu\nu}\, \omega \log \frac{\omega}{M}+\dots\,,\\
I_3^{\mu\nu}&=(p_i\cdot q)^2\,\bar I_3^{\mu\nu}\,\omega^2\log\frac{\omega}{M}+\dots\,,
\end{alignedat}
\label{logCoefficients}
\end{equation}
where the coefficients $\bar I_1$, $\bar I_2^\mu$, $\bar I_3^{\mu\nu}$, $\bar J_2^\mu$, $\bar J_3^{\mu\nu}$ depend only on $p_i$ and $p_j$, and the dots denote terms without $\log\omega$.

We apply two operations in order to select the relevant terms. First, $L_\omega$ keeps only the coefficient of $\omega^0\log\omega$, so terms of order $\omega\log\omega$ drop out. Second, $D^2$ annihilates any term independent of the soft direction $q$. 
Inserting \eqref{logCoefficients} in the amplitude, the $I_1$ term and the anticommutator terms multiplying $I_2^\mu$ and $J_2^\mu$ are independent of $q$, while the remaining terms with $I_2^\mu$, $I_3^{\mu\nu}$ and $J_3^{\mu\nu}$ are of order $\omega\log\omega$.
The only surviving contribution is the $k_\mu J_2^\mu$ term, in which $k_\mu/(p_i\cdot k)$ is of order $\omega^0$ and depends on $q$:
\be
D^2L_\omega[\cA_a+\cA_b]
=-g^3\,D^2\!\left[\frac{8M^3}{p_i\cdot q}\,\big(q\cdot\bar J_2\big)\right]\bar u(p_i)\,H\,u(p_j)\,.
\ee
By Lorentz covariance, $\bar J_2^\mu$ can only be built from $p_i^\mu$ and $p_j^\mu$. In the previous expression, a component along $p_i^\mu$ gives $q\cdot p_i/(p_i\cdot q)=1$, which is independent of $q$ and is annihilated by $D^2$. Evaluating the integral, we find
\be
\bar J_2^\mu=\frac{i}{64\pi^2}\,\frac{\partial I^{\rm q}_{ij}}{\partial p_{i\mu}}+( \cdots )\,p_i^\mu\,,
\ee
with $I^{\rm q}_{ij}$ defined in \eqref{Iq}. Since $I^{\rm q}_{ij}$ depends on $p_i$ only through $p_i\cdot p_j$, the first term points along $p_j^\mu$. Therefore,
\be
D^2L_\omega[\cA_a+\cA_b]
=-\frac{ig^3M^3}{8\pi^2}\,D^2\!\left[\frac{1}{p_i\cdot q}\left(q\cdot\frac{\partial}{\partial p_i}\right)I^{\rm q}_{ij}\right]\bar u(p_i)\,H\,u(p_j)\,.
\ee

\section{Renormalization of Yukawa theory and single-legs contributions}
\label{AppRen}

Ultraviolet divergences are regulated in $d=4-2\varepsilon$ dimensions, with $g\to g\bar\mu_R^{\varepsilon}$ and the usual constants absorbed in the renormalization scale $\mu_R$, which we distinguish from the sphere smearing $\mu$. A small scalar mass $m_{\rm IR}$ is used only as an infrared regulator, and is removed after the regulated combinations have been formed. The counterterms are
\begin{align}
\cL_{\rm ct}=&\,(Z_\Psi-1)\,i\bar\Psi\slashed\d\Psi-(Z_M-1)M\bar\Psi\Psi-\frac{1}{2}(Z_\phi-1)(\d\phi)^2\nonumber\\
&+g(Z_g-1)\phi\bar\Psi\Psi+\frac{1}{2}Z_mM^2\phi^2+\cL^{\rm local}_{\rm ct}\,.
\end{align}
The last term contains the scalar tadpole and the $\phi^3$, $\phi^4$ counterterms discussed in Section~\ref{sec:setup}. Their subtraction conditions specify the scalar background about which the Yukawa calculation is performed. The expressions below concern the two-point and Yukawa-vertex sectors.

\subsection*{Scalar propagator}

For the closed fermion loop,
\be
\tr\Big[(-\slashed\ell-\slashed k+M)(-\slashed\ell+M)\Big]=-4(\ell+k)\cdot\ell+4M^2\,.
\ee
The loop is infrared safe. With a Feynman parameter and $q=\ell+xk$,
\be
\frac{1}{[(\ell+k)^2+M^2](\ell^2+M^2)}=\int_0^1\frac{dx}{(q^2+D_\phi)^2}\,,\qquad D_\phi=M^2+x(1-x)k^2\,.
\ee
The denominator remains squared after the shift, and the term odd in $q$ integrates to zero. Using
\be
\bar\mu_R^{2\varepsilon}\int\frac{d^dq}{(2\pi)^d}\frac{1}{(q^2+D)^2}=\frac{i}{16\pi^2}\left(\frac{1}{\varepsilon}-\log\frac{D}{\mu_R^2}\right),\qquad
\bar\mu_R^{2\varepsilon}\int\frac{d^dq}{(2\pi)^d}\frac{q^2}{(q^2+D)^2}=-\frac{2iD}{16\pi^2}\left(\frac{1}{\varepsilon}+\frac{1}{2}-\log\frac{D}{\mu_R^2}\right),
\ee
the loop gives
\be
\Pi_{\phi,{\rm loop}}(k^2)=-\frac{g^2}{4\pi^2}\left[\left(M^2+\frac{k^2}{6}\right)\left(1+\frac{3}{\varepsilon}\right)-3\int_0^1dx\,D_\phi\log\frac{D_\phi}{\mu_R^2}\right].
\ee
The counterterm contribution is $\Pi_{\phi,{\rm ct}}(k^2)=-(Z_\phi-1)k^2-Z_mM^2$, and the conditions \eqref{renCond} fix
\be
Z_m=-\frac{g^2}{4\pi^2}\left[\frac{3}{\varepsilon}+1-3\log\frac{M^2}{\mu_R^2}\right]+\cO(g^4)\,,\qquad
Z_\phi=1-\frac{g^2}{24\pi^2}\left[\frac{3}{\varepsilon}-2-3\log\frac{M^2}{\mu_R^2}\right]+\cO(g^4)\,.
\ee
The renormalized two-point function is
\be
\Pi^{\rm ren}_\phi(k^2)=\frac{g^2}{4\pi^2}\left[-\frac{k^2}{2}+3\int_0^1dx\,D_\phi\log\frac{D_\phi}{M^2}\right],
\ee
whose value and first derivative vanish at $k^2=0$, as required.

\subsection*{Fermion propagator}

Define
\be
D_\Psi(x,p)=x(1-x)p^2+x\,m_{\rm IR}^2+(1-x)M^2\,,\qquad \bar D_\Psi(x)=(1-x)^2M^2+x\,m^2_{\rm IR}\,.
\ee
The one-loop self-energy is
\be
\Pi_{\Psi,{\rm loop}}(\slashed p)=\frac{g^2}{16\pi^2}\int_0^1dx\,(-x\slashed p+M)\left[\frac{1}{\varepsilon}-\log\frac{D_\Psi}{\mu_R^2}\right].
\ee
We parametrize the counterterms by
\be
Z_\Psi=1+\frac{g^2}{16\pi^2}\left(-\frac{1}{2\varepsilon}+\bar Z_\Psi\right),\qquad
Z_M=1+\frac{g^2}{16\pi^2}\left(\frac{1}{\varepsilon}+\bar Z_M\right).
\ee
The on-shell conditions at $\slashed p=-M$ give
\be
\bar Z_\Psi=\int_0^1dx\,x\log\frac{\bar D_\Psi}{\mu_R^2}-I_0\,,\qquad
\bar Z_M=-\int_0^1dx\log\frac{\bar D_\Psi}{\mu_R^2}-I_0\,,\qquad
I_0=\int_0^1dx\,\frac{2M^2x(1-x^2)}{\bar D_\Psi}\,,
\ee
and the renormalized self-energy is
\be
\Pi^{\rm ren}_\Psi=\frac{g^2}{16\pi^2}\left[-\int_0^1dx\,(-x\slashed p+M)\log\frac{D_\Psi}{\bar D_\Psi}+(\slashed p+M)I_0\right].
\ee
Ultraviolet finiteness does not imply infrared finiteness: $I_0$ contains a logarithm as $m_{\rm IR}\to0$.

\subsection*{The Yukawa vertex}

For a soft scalar of momentum $k$, the loop vertex is
\be
i\Pi_{V,{\rm loop}}(p,k)=(ig)^3\left(\frac{1}{i}\right)^3\int\frac{d^d\ell}{(2\pi)^d}\frac{[-(\slashed p+\slashed k+\slashed\ell)+M][-(\slashed p+\slashed\ell)+M]}{(\ell^2+m^2_{\rm IR})[(p+k+\ell)^2+M^2][(p+\ell)^2+M^2]}\,.
\ee
We define the simplex measure and denominator by
\begin{gather}
\int dF_3\,f(x,y,z)=2\int_0^1dx\,dy\,dz\,\delta(x+y+z-1)f(x,y,z)\,,\\
\tilde D=x(1-x)p^2+y(1-y)k^2+2xy\,p\cdot k+(1-x)M^2+x\,m^2_{\rm IR}\,.
\end{gather}
After shifting $\ell=q-(1-x)p-yk$, the odd terms in $q$ vanish. The ultraviolet pole is removed by
\be
Z_g=1+\frac{g^2}{16\pi^2}\left(\frac{1}{\varepsilon}+\bar Z_g\right),
\ee
and the tree vertex $g$ is not included again in the counterterm. For an on-shell external spinor with $p=p_i$, introduce
\be
s=\frac{2p_i\cdot k}{M^2}\,,\qquad
\bar u(p_i)\Pi^{\rm ren}_V=\frac{g^3}{16\pi^2}\bar u(p_i)\left[\Pi_0(s)+\frac{\slashed k}{M}\Pi_k(s)\right],
\ee
with, after removal of the infrared regulator at nonzero $s$,
\begin{align}
\Pi_0(s)&=\bar Z_g+\frac{1}{2}-\log\frac{\mu_R^2}{M^2}+J_0(s)+J_1(s)\,,\qquad \Pi_k(s)=J_2(s)\,,\\
J_0(s)&=\int dF_3\log\Big[(1-x)^2+xys\Big]\,,\\
J_1(s)&=\int dF_3\,\frac{(1+x)^2-xys}{2[(1-x)^2+xys]}\,,\\
J_2(s)&=\int dF_3\,\frac{2y-1-x}{2[(1-x)^2+xys]}\,.
\end{align}
These formulas may be evaluated for positive $s$ and then continued with the Feynman prescription. On physical outgoing momenta in the mostly-plus convention, $p_i\cdot k$ is negative, so the continuation must be retained when combining phases.

\subsection*{Soft emission from a single external line}

Let $p=p_i+k$ and let $\cH$ be the amputated hard spinor amplitude. The two insertions are
\begin{align}
S_{\rm prop}&=\bar u(p_i)\,ig\left(\frac{1}{i}\right)^2\frac{-\slashed p+M}{p^2+M^2}\,i\Pi^{\rm ren}_\Psi(\slashed p)\,\frac{-\slashed p+M}{p^2+M^2}\,\cH\,,\\
S_{\rm vertex}&=\bar u(p_i)\,i\Pi^{\rm ren}_V(p_i,k)\,\frac{1}{i}\,\frac{-\slashed p_i-\slashed k+M}{2p_i\cdot k}\,\cH\,.
\end{align}
The self-energy can be written as
\be
\Pi^{\rm ren}_\Psi=\frac{g^2}{16\pi^2}\Big[\slashed p\,I_2-M\,I_1+(\slashed p+M)I_0\Big]\,,\qquad
I_1=\int_0^1dx\log\frac{D_\Psi}{\bar D_\Psi}\,,\quad I_2=\int_0^1dx\,x\log\frac{D_\Psi}{\bar D_\Psi}\,.
\ee
On the configuration of interest $D_\Psi=\bar D_\Psi+x(1-x)M^2s$, and at $m_{\rm IR}=0$ and nonzero $s$,
\be
I_1=-\frac{s}{1-s}\log s\,,\qquad I_2=-\frac{s}{2(1-s)}+\frac{1}{2}\log s-\frac{\log s}{2(1-s)^2}\,.
\ee
The regulator must remain in $I_0$. Since the propagator insertion contains two nearly on-shell propagators, $I_1$ and $I_2$ must be expanded through $s^2\log s$ to obtain the logarithmic term of order $\omega^0$. The result is
\be
S_{\rm prop}=\frac{g^3}{16\pi^2M}\bar u(p_i)\left[\frac{2}{s}\big(I_0+1+4\log s\big)+\frac{1}{2}+6\log s-\frac{\slashed k}{Ms}\big(I_0+1+4\log s\big)\right]\cH+\cO(s\log s)\,.
\ee
The vertex integrals needed at the same accuracy are
\be
J_0(s)=-1+\frac{s}{2}+\cO(s^2\log s)\,,\qquad
J_1(s)=\frac{1}{2}-\frac{s}{2}-4(1+s)\log s+\cO(s^2\log s)\,,\qquad
J_2(s)=2\log s+\cO(s\log s)\,.
\ee
The analytic term $-s/2$ in $J_1$ cancels the corresponding term in $J_0$; it is irrelevant to the logarithmic cancellation but matters if a finite $\omega^0$ term is retained. Defining $c_g=\bar Z_g-\log(\mu_R^2/M^2)$, the vertex insertion becomes
\be
S_{\rm vertex}=\frac{g^3}{16\pi^2M}\bar u(p_i)\left[\frac{2}{s}\big(c_g-4\log s\big)-6\log s-\frac{\slashed k}{Ms}\big(c_g-4\log s\big)\right]\cH+\cO(s\log s)\,.
\ee
The terms $\log s/s$, $\log s$ and $\slashed k\log s/s$ cancel separately in the sum:
\be
S_{\rm prop}+S_{\rm vertex}=\frac{g^3}{16\pi^2M}\bar u(p_i)\left[\frac{2}{s}\left(1-\frac{\slashed k}{2M}\right)\big(I_0+1+c_g\big)+\frac{1}{2}\right]\cH+\cO(s\log s)\,.
\label{Stotal}
\ee
This cancellation does not require a choice of $\bar Z_g$. A convenient soft-pole normalization is
\be
\bar Z_g=\log\frac{\mu_R^2}{M^2}-I_0-1\,,
\label{Zgbar-fix}
\ee
which absorbs the remaining single-line pole into the definition of the renormalized Yukawa coupling at fixed infrared regulator. Its ultraviolet-finite part contains the infrared logarithm in $I_0$, as can occur in on-shell subtraction prescriptions; see section 18.3.1 of \cite{Schwartz:2014sze}. This should not be confused with an infrared-finite definition of every scattering observable; the genuine infrared structure of the process resides in the two-line kernels $I_{ij}$. The finite remainder in \eqref{Stotal} belongs to this particular set of diagrams and subtraction conditions and does not establish a scheme-independent subleading soft theorem. The main text uses only the absence of a surviving logarithm from the single-line sum.

\section{The retarded field of an incoming particle}
\label{AppCoulomb}
We now verify the antipodal matching relation exhibited in \eqref{eq:fut} and  \eqref{eq:pas} directly from the retarded field generated by the asymptotic trajectories. We consider a single incoming particle and compute its contribution to the $1/u$ tail at $\dI^+_-$, the $\eta^{-2}\log\eta$ coefficient near $i^0$, and the $r^{-2}\log r$ coefficient at $\dI^-_+$. Since the scalar field equation is linear, the result for a collection of incoming particles follows by summing their individual contributions.

The calculation identifies the spatial infinity coefficients $\Lambda_{\ell m}$ with the spherical harmonics of $\chi^{+,\rm pp}_-$ defined in
\eqref{retardedEndpointCoefficients} for $\ell\geq1$. 

The asymptotic trajectory is then
\begin{equation}
X^\mu(\tau)=\alpha^\mu+V^\mu\tau
+C^\mu\log\frac{-\tau}{\tau_0}+\cdots,
\qquad \tau\to-\infty,
\end{equation}
where $V^2=-1$ and $C^\mu$ describes the logarithmic correction induced by the long-range scalar interaction.  It is convenient to introduce
\begin{equation}
K=\frac{\kappa}{4\pi}\,,\qquad
C_\perp^\mu=C^\mu+(V\cdot C)V^\mu\,,\qquad
V\cdot C_\perp=0\,,
\end{equation}
where $C_\perp^\mu$ is the component of the logarithmic correction orthogonal to the asymptotic velocity. This decomposition will be useful in identifying the logarithmic coefficients of the field. Using the retarded solution \eqref{phi}, the scalar field generated by this particle can be written as
\begin{equation}
\Phi(x)=\frac{K}{\rho}\,,\qquad
\rho=-\dot X(\tau^*)\cdot\bigl(x-X(\tau^*)\bigr)\,,
\label{rhoDef}
\end{equation}
where $\tau^*$ is the retarded root of $(x-X(\tau))^2=0$. In the following expansions, we retain the coefficients relevant for the logarithmic matching. The constant displacement $\alpha^\mu$ is omitted from the contribution presented here.

\subsection*{Future null infinity}
We first examine the behavior of the retarded field at future null infinity. At fixed $u$ and large $r$, the leading null condition $(x-X(\tau^*))^2=0$ reduces to
$q\cdot X(\tau^*)=-u$, with $q^\mu=(1,\hat x)$. Since
$q\cdot V<0$, the early-time root is
$\tau^*=u/(-q\cdot V)+O(\log|u|)$ as $u\to-\infty$.
The leading radial coefficient of \eqref{rhoDef} is then
\begin{equation}
\varphi_R(u,\hat x)
=
-\frac{K}{q\cdot\dot X(\tau^*)}
=
-\frac{K}{q\cdot V}
+\frac{\chi^{+,\rm pp}_-(\hat x)}{u}
+O\!\left(\frac{\log|u|}{u^2}\right),
\qquad
\chi^{+,\rm pp}_-(\hat x)=-K\,\frac{q\cdot C}{q\cdot V}\,.
\label{retardedTailSingleParticle}
\end{equation}
This is the single particle contribution to \eqref{retardedEndpointCoefficients}. The
$1/u$ tail comes from the velocity correction $C^\mu/\tau^*$. At fixed $u$, the retarded root stays finite as $r\to\infty$, so the displacement $C^\mu\log(-\tau^*/\tau_0)$ introduces no radial logarithm and $\Phi^+_-=0$.

\subsection*{Spatial infinity}

To parameterize spatial sections we use $x^\mu=\eta N^\mu$ with $N^\mu=(s,\hat x)/\sqrt{1-s^2}$ and $N^2=1$. At fixed $s$, the leading retarded root is
\begin{equation}
\tau^*=\eta T+O(\log\eta)\,,\qquad
T=-N\cdot V-\sqrt{1+(N\cdot V)^2}\,.
\end{equation}
Thus $\log(-\tau^*/\tau_0)=\log(\eta/\tau_0)+O(1)$.
The correction $C^\mu/\tau^*$ to the velocity is of order
$\eta^{-1}$ and contributes at order $\eta^{-2}$ without a
logarithm. 

To extract the logarithmic term, we can therefore
replace the trajectory by the straight line
$X^\mu=V^\mu\tau+LC^\mu$ with $L=\log(\eta/\tau_0)$ held fixed
when solving the null condition. Its exact retarded distance is
\begin{equation}
\rho_L
=
\sqrt{(x-LC)^2+\bigl[V\cdot(x-LC)\bigr]^2}\,.
\label{rhoDisplacedLine}
\end{equation}
Expanding \eqref{rhoDisplacedLine} at fixed $s$ gives
\begin{equation}
\Phi(\eta,s,\hat x)
=
\frac{K}{\eta\sqrt{1+(N\cdot V)^2}}
+\frac{\log(\eta/\tau_0)}{\eta^2}\,\Lambda^{(1)}(s,\hat x)
+O(\eta^{-2})\,,
\qquad
\Lambda^{(1)}(s,\hat x)
=
K\,\frac{N\cdot C_\perp}{\bigl[1+(N\cdot V)^2\bigr]^{3/2}}\,.
\label{LambdaParticle}
\end{equation}
To identify the coefficients in the expansion \eqref{ThetaLogEta}, we examine the behavior of $\Lambda^{(1)}$ near $s=1$. Since
$\sqrt{1-s^2}\,N^\mu\to q^\mu$ as $s\to1^-$, we find
\begin{equation}
\lim_{s\to1^-}\frac{\Lambda^{(1)}(s,\hat x)}{1-s^2}
=
K\,\frac{q\cdot C_\perp}{(-q\cdot V)^3}
=
-\frac12D^2\chi^{+,\rm pp}_-(\hat x)\,.
\label{LambdaParticleBoundary}
\end{equation}
The last equality uses the spherical identity
\begin{equation}
D^2\!\left(\frac{q\cdot C}{q\cdot V}\right)
=
-\frac{2\,q\cdot C_\perp}{(q\cdot V)^3}\,,
\label{ParticleSphereIdentity}
\end{equation}
which follows by the product rule from
$D^2(\hat x\cdot\vec b)=-2\hat x\cdot\vec b$ and
$D_A(\hat x\cdot\vec b)D^A(\hat x\cdot\vec c)
=\vec b\cdot\vec c-(\hat x\cdot\vec b)(\hat x\cdot\vec c)$
for constant vectors $\vec b$ and $\vec c$, together with
$V^2=-1$.

\noindent In the $P$-branch normalization of \eqref{ThetaQ1}, the same
limit computed from \eqref{ThetaLogEta} is
\begin{equation}
\lim_{s\to1^-}\frac{\Lambda^{(1)}(s,\hat x)}{1-s^2}
=
\sum_{\ell\geq1,m}\frac{\ell(\ell+1)}{2}\,
\Lambda_{\ell m}Y_{\ell m}(\hat x)
=
-\frac12D^2\Lambda(\hat x)\,,
\end{equation}
where we used $P'_\ell(1)=\ell(\ell+1)/2$. Comparing with
\eqref{LambdaParticleBoundary} gives
\begin{equation}
\Lambda_{\ell m}=(\chi^{+,\rm pp}_-)_{\ell m}\quad(\ell\geq1)\,.
\label{ParticleBranchCoefficients}
\end{equation}
The projection removes the $\ell=0$ mode, including the constant
contribution $K \, V \cdot C$ from the part of $C^\mu$ parallel
to $V^\mu$ in \eqref{retardedTailSingleParticle}.

\subsection*{Past null infinity}

We finally study the retarded field at past null infinity. At fixed advanced time $v=t+r$, we write $x^\mu=r\bar q^\mu+v e^\mu$, with
$\bar q^\mu=(-1,\hat x)$ and time-like vector $e^\mu=(1,\vec 0)$.
Since $\bar q\cdot V>0$, the leading retarded root behaves as $\tau^*=-2r\,\bar q\cdot V+O(\log r)$.
The logarithmic correction to the root is determined by the null condition and implies $\log(-\tau^*/\tau_0)=\log(r/\tau_0)+O(1)$. To obtain the asymptotic field, we evaluate the retarded distance $\rho$.
The velocity correction is of order $r^{-1}$ and gives a nonlogarithmic term of order $r^{-2}$ in the field. The
logarithmic coefficient again follows from
\eqref{rhoDisplacedLine}, now with,
\begin{equation}
\Phi(v,r,\hat x)
=
\frac{K}{r\,\bar q\cdot V}
+\frac{\log(r/\tau_0)}{r^2}\,\Phi^-_+(\hat x)
+O(r^{-2})\,,
\qquad
\Phi^-_+(\hat x)
=
K\,\frac{\bar q\cdot C_\perp}{(\bar q\cdot V)^3}\,.
\label{coulombLog}
\end{equation}
The leading coefficient $K/(\bar q\cdot V)$ is independent of
$v$, so $\chi^-_+=0$ for this retarded solution.

\subsection*{Antipodal matching}
Using $\bar q^\mu(-\hat x)=-q^\mu(\hat x)$ in
\eqref{coulombLog} and applying \eqref{ParticleSphereIdentity}, we find
\begin{equation}
\Phi^-_+(-\hat x)
=
-K\,\frac{q\cdot C_\perp}{(-q\cdot V)^3}
=
\frac12D^2\chi^{+,\rm pp}_-(\hat x)
=
\frac12D^2\Lambda(\hat x)\,.
\end{equation}
In the rest frame, this reduces to the dipole relation
\begin{equation}
\Phi^-_+(-\hat x)
=
-K\,\hat x\cdot\vec C
=
\frac12D^2\bigl(K\,\hat x\cdot\vec C\bigr)\,,
\qquad
\Lambda^{(1)}(s,\hat x)
=
K\,(\hat x\cdot\vec C)(1-s^2)\,.
\end{equation}
Together with $\Phi^+_-=\chi^-_+=0$, the antipodal relation is verified for the retarded solution.


\providecommand{\href}[2]{#2}\begingroup\raggedright\endgroup

\end{document}